\documentclass[aps,amsmath,nofootinbib,superscriptaddress]{revtex4}

\usepackage{psfig}
\usepackage{graphicx}
\usepackage[latin1]{inputenc}

\usepackage{color} 

\definecolor{Trot}{rgb}{0.0, 0.0, 1.0}     
\definecolor{joerg}{rgb}{1.0,0.0,0.0}

\newcommand{\nn}{\nonumber}
\newcommand{\be}{\begin{equation}}
\newcommand{\bel}[1]{\begin{equation} \label{#1} }
\newcommand{\ee}{\end{equation}}
\newcommand{\bea}{\begin{eqnarray}}
\newcommand{\beal}[1]{\begin{eqnarray} \label{#1} }
\newcommand{\eea}{\end{eqnarray}}
\renewcommand{\d}{{\rm d}}

\begin{document}

\title{Non-perturbative finite $T$ broadening of the $\rho$ meson and dilepton emission in heavy ion-collisions}

\author{J\"{o}rg Ruppert}
\email{ruppert@phy.duke.edu} 
\author{Thorsten Renk}
\email{trenk@phy.duke.edu}
\affiliation{
 Department of Physics,\\
 Duke University,
 Physics Bldg., Science Drive,
 Box 90305,\\
 Durham, 27708-0305 NC, USA}
\begin{abstract}
\vspace{1cm}
\end{abstract}

\begin{abstract}
We study self-consistently the finite $T$ broadening of the $\rho$-meson and its implications for dilepton emission in heavy ion-collisions. For this purpose finite width effects at finite temperature due to the $\rho$-$\pi$ interaction are investigated in a self-consistent $\Phi$-functional approach. The temperature dependence of the $\rho$-meson and pion spectral functions and self-energies is discussed. The spectral functions show considerable broadening in comparison with a perturbative calculation on the one-loop level. Using these spectral functions, we make a comparison to dilepton emission data from the CERES NA49 collaboration employing a parametrized fireball evolution model of collision. We demonstrate that these non-perturbative finite width effects are in-medium modifications relevant to the understanding of the enhancement of the low invariant mass spectrum of dileptons emitted in A-A collisions. 
\end{abstract}

\maketitle
\section{Introduction}
\label{Introduction}

In recent years many theoretical and experimental research has
been undertaken to investigate the physics of ultra-relativistic
heavy-ion collisions (URHIC). The important goal here
is to gain a better understanding of the properties of extremely hot and dense nuclear
matter. Through
variation of the experimental parameters - e.g.
collision energy and impact parameter - different regimes of temperature and net-baryon density of the nuclear system are
explored. The ultimate goal is to understand the phase
structure of quantum chromodynamics (QCD), the theory of strong interactions.

In order to extract reliable information about the properties of
nuclear matter under these extreme conditions 
(and especially to decide if a
phase transition from the phase of (confined) hadronic matter to the
(partonic) quark-gluon-plasma (QGP) has taken place) one has to do a careful and
comprehensive analysis of the experimentally accessible observables.
Dileptons ($e^+ e^-$ and $\mu^+
\mu^-$ pairs) and real photons are interesting probes
in this context. 
\noindent
Since they are emitted during all collision phases
and leave the interaction region without substantial rescattering due to
the smallness of the electromagnetic coupling.  Hence, they reflect the
whole space-time evolution of the collision process, in contrast to hadronic
probes which predominantly reflect the conditions at kinetic decoupling.

To some degree, measurements made so far complement each other:
while photons have been measured for comparative large transverse momentum scales of
1.5 GeV and above, the invariant mass spectrum of dielectrons is rather well
known below 1 GeV. Therefore, the measured photons naturally carry
information about the early, hard partonic stages of the collision process
whereas the scale set by the dilepton measurement dominantly reflect the later, soft  hadronic stages.

While there is little reason to assume that low mass dilepton physics
would be dominated by emission from a QGP, understanding the
late hadronic phase of the evolution is nevertheless of crucial
importance for forming a comprehensive picture of the whole collision process.
The most intriguing physics question here is the possibility
of in-medium modifications of hadron properties. The CERES and HELIOS-3
collaborations demonstrated that central nucleus-nucleus (A-A)
collisions exhibit a strong enhancement of low-mass dileptons in
the range between $200 \text{ MeV}$ and $600 \text{ MeV}$
 as compared to the yield
one expects from the corresponding rates of proton-nucleus (p-A)
and proton-proton collisions \cite{Masera:1995ck,Agakishiev:1995xb,
Agakishiev:1997au,Lenkeit:1999xu,Porter:1997rc,Wilson:1997sr}
which may hint at such modifications. For an
overview of the various mechanisms brought forward to explain this
enhancement see e.g. \cite{Rapp:1999ej}. 

It is often argued that a large broadening of
the $\rho$-meson spectral function cannot be expected by a purely mesonic medium. This assumption is based on perturbative calculations of finite width effects of pions on the $\rho$-meson at finite temperature \cite{Gale:1991pn}. However, these investigations consider the width of the $\rho$-meson caused by stable particles only.
In most of the theoretical investigations the damping width due to collisional broadening 
is ignored or investigated in an extended perturbative framework\cite{Urban:1998eg}. Although investigation of collisional broadening effects due to the interaction of vector
mesons with thermal mesons in a kinetic theory framework have been carried out, they consider possible scattering processes of {\it on-shell} mesons only \cite{Haglin:1994xu,Gao:1998mn}.  In contrast we extend these studies in order to include also {\it off-shell} contributions.

In the present paper we study these broadening effects and demonstrate that even the $\pi$ - $\rho$ interaction alone is able to induce substantial broadening of the $\rho$-mesons's spectral function  due to the in-medium damping width of the pion.  

For the purpose of studying this mechanism, we employ a many body resummation which is known as the $\Phi$-functional approach. In this approach we derive Dyson Schwinger equations for the propagators
of the $\rho$-meson and the pion. Our calculation are {\it self-consistent} in the sense 
that the in-medium broadening of the $\rho$-meson and the pions is taken taken into account by this scheme in its reciprocal dependence.

We concentrate on the {\it phenomenology} of collisional broadening and leave the effect of mass shifts relative to the vaccum mass of the particles (especially problems of renormalization) aside by taking into account only the imaginary parts of the self-energies.

Since our primary aim in this study is to investigate how the $\pi$ - $\rho$ interaction, giving raise to the dominant decay channel of the $\rho$-meson in the vacuum, is modified at finite temperature, we disregard other effects in our model (like mass shifts, scattering off by baryons and effects of chiral symmetry restoration) focussing on collisional broadening due to that interaction only. For approaches considering these other effects, see e.g.  \cite{Rapp:1999ej,Renk:2002md,Renk:2003hu}. 

We compare the results of this self-consistent calculation with a perturbative analysis and show that the self-consistent treatment of this interaction leads to considerable collisional in-medium broadening at finite temperature.

In order to demonstrate that the broadening effects are quantitatively important for the understanding of the observed low mass dilepton enhancement as seen in experiments, we  compare the model predicitons with the experimental results obtained by the CERES-collaboration 
\cite{Agakishiev:1995xb,Agakishiev:1997au,Lenkeit:1999xu}. 
For this purpose, we fold the predicted rate with a  parametrized fireball evolution
which has been shown to successfully reproduce a number of different observables, among them charmonium suppression, direct photon emission, hadronic transverse momentum spectra and Hanbury Brown Twiss correlation measurements \cite{Renk:2004cj}.

The paper is organized as follows: We first introduce the
framework for the description of the $\pi$-$\rho$ interaction.
Then, the self-consistent resummation scheme is described and
the approximations of the Schwinger-Dyson
equations for the $\rho$-meson and the pion are derived where we
employ the St\"uckelberg method to describe the $\rho$-mesons as
massive vector mesons.   
Special care has to be taken to propagate only
four-dimensionally transverse and therefore physical modes of the
vector-mesons in order not to violate current conservation. 
Two different solutions of these difficulties are discussed. 
In section \ref{width} we present the
solution of the self-consistent equations for the spectral
functions of the pions and the $\rho^0$-meson and discuss the
quantitative and qualitative features of the in-medium
modifications which are important in the hadronic phase.
Section \ref{QGP} contains a brief
review of the model description of the QGP
phase, section \ref{hadronic} deals with non-thermal contributions to the
dilepton yield, section \ref{fireball} provides a short introduction to the
parametrization of the fireball evolution.
Finally, in section \ref{massspectrum} we compare the model calculation with
the dilepton spectra as obtained by the CERES collaboration and
discuss these results. We end with a conclusion and an outlook.

\section{The $\rho$-meson $\pi$ interaction model}
\label{interaction}

We restrict our self-consistent study of the in-medium properties
of the $\rho$-meson to the $\rho$-$\pi$ interaction,
emphasizing self-consistent finite pion width effects on the
$\rho$-meson. The framework of these considerations is a model
inspired by vector meson dominance (VMD) \cite{Sakurai:1960ju,Kroll:1967it} 
for the interaction of neutral $\rho$-mesons with the charged pions.

In order to describe the propagation of the massive vector mesons
in the {\it appropiate} quantum field theoretical framework, in order
to propagate their physical modes, we treat the $\rho$-meson in
the St\"uckelberg formalism \cite{Stueckel:1939ii}. 
(For a detailed review of this 
formalism see also \cite{Ruegg:2003ps}.)
The suitability of this formalism for calculations in $\Phi$-functional approaches
was first discussed in \cite{HendrikDiss}.
In the St\"uckelberg framework one introduces an additional scalar field, the so called
Stückelberg ghost $\phi$. The free Langrangian density of the
$\rho$-meson field $A_\mu$  and the Stückelberg ghost $\phi$ is given by:
\bea {\cal L}_{\rm Stueck}=-\frac{1}{4}{\cal F}_{\mu \nu}{\cal F}^{\mu
\nu}+\frac{m^2}{2}A_{\mu}A^{\mu} +\frac{1}{2}(\partial_\mu
\phi)(\partial^\mu \phi) + m \phi \partial_\mu A^\mu\,\,,\eea
where  ${\cal F}_{\mu, \nu}=\partial_\mu A_\mu - \partial_\nu A_\mu$ is the 
corresponding field strength tensor.
Pauli showed first \cite{Pauli:1941} 
that this treatment of massive vector mesons
leads to a symmetry of the action (up to a total divergence) under
the following (local) transformations of the fields:. \bea A_\mu
&\rightarrow&
A'_\mu=A_\mu+\delta A_\mu=A_\mu+\partial_\mu \lambda \, \, ,\nn \\
\phi &\rightarrow& \phi'=\phi+\delta \phi=\phi+m\lambda \, \,.\eea

This gauge symmetry has to be taken into account when quantizing
the theory. Applying the standard Fadeev Popov quantization scheme
and introducing a gauge fixing term in the $R_\xi$-gauge \bea
\label{gaugefix} g(A,\phi)=\partial_\mu A^\mu + \xi m \phi \,\,
\eea one obtains after quantization the following effective free
Lagrangian density including Stückelberg ghosts $\phi$ and
Fadeev Popov ghosts $\eta$:
\bea \label{effLagPiRho}{\cal L}_{\rm
eff}&=&-\frac{1}{4}{\cal F}_{\mu \nu}{\cal F}^{\mu
\nu}+\frac{1}{2}m^2\left(A^\mu-\frac{1}{m}\partial^\mu
\phi\right)^2-\frac{1}{2\xi}\left(\partial_\mu A^\mu+\xi m \phi\right)^2 \nn \\
&& -\eta^{*}\left(\partial^2+\xi m^2\right)\eta \, \,. \eea

Expanding the contributions in the effective free Lagrangian
density and partial integration shows that in the $R_\xi$-gauge
the fields $A_\mu$, $\phi$ and $\eta$ all decouple from each other.

There are many advantages of this description of massive vector
mesons. The theory is symmetric under the appropriate
Becchi-Rouet-Stora-Tyutin (BRST) - transformations \cite{Becchi:1974xu,Becchi:1975md,Tyutin:1975qk} and leads to a
positive definite Hamiltonian for the physical propagating 
modes \cite{Ruegg:2003ps}.
Moreover this leads to proof of the unitarity and can be used to
demonstrate (perturbative) renormalizability \cite{Ruegg:2003ps,vanHees:2003dk}.

The neutral $\rho$-mesons are in a
VMD-inspired model decisive for the dilepton
production, which is given in the hot $\rho-\pi$ medium by
the decay process $\rho^0 \rightarrow \gamma^* \rightarrow e^+
+e^-$  via an intermediate virtual photon $\gamma^*$.
We now include the interaction of the neutral
$\rho$-mesons with the charged pions. 

The pion is coupled minimally to the vector meson field $A_\mu$ and to
the electromagnetic photon field $A_{\mu,\gamma}$. Therefore the part of the
Lagrangian density including these interactions is given by:
 \bea \label{Pionterm2}{\cal L}_{\pi}&=&
\left[\left(\partial_\mu + i g A_\mu + i e A_{\mu,{\rm
Photo}}\right)\pi\right]^{*} \left[\left(\partial^\mu + i g A^\mu+
i e A^{\mu,\gamma}\right)\pi\right] \nn \\ &&-m_{\pi}^2
\pi^{*} \pi
 \, \,. \eea
Additionally, we introduce an interaction term coupling the photon
to the $\rho$-meson vector field: \bea \label{Photo}
   {\cal L}_{\rho - \gamma}=-\frac{e}{2g}F_{\mu \nu}{\cal F}^{\mu
   \nu} \, \, ,
 \eea where ${\cal F}^{\mu \nu}$ and $F^{\mu \nu}$ are the 
field strength tensors of the rho-meson and photon, respectively.

The total Lagrangian density \footnote{It should be noted, that if
one neglects all terms including the $A_{\gamma}$ field, this
Lagrangian for the pions and the $\rho^0$-meson is formally
identical to that one for scalar quantum electrodynamics with a
massive photon. The pions as pseudoscalar have taken the role of
the scalar particle and the $\rho^0$-meson has taken the formal
role of a massive photon.} is now the sum of the three
contributions discussed above:\bea\label{effLagPiRho2} {\cal
L}_{\pi - \rho} = {\cal L}_{\rm eff}+{\cal
L}_{\rho - \gamma}+{\cal L}_{\pi}\,\,. \eea

The electromagnet current $J^h_\mu$ created by the mesons is given
by: \bea \label{StromEM} J_{\nu}^{\rm
h}=\frac{e^2}{g^2}\partial_{\mu}F^{\mu}_{\nu} +
\frac{e}{g} m^2 A_\nu \, \,  \eea and is in order $e/g$
proportional to the $\rho$ field. This is the manifestation of
VMD.

The coupling of the photons-field to dileptons can be calculated in quantum
electrodynamics (QED). This is essential for describing the
production of dileptons from the hot hadronic medium in VMD, see appendix
\ref{production}.

\section{Self-consistent resummation scheme}
\label{scheme}

We employ a self-consistent many body resummation scheme which has
been first introduced by Baym \cite{Baym:1962sx} and is known as the
$\Phi$-functional formalism to model self-consistently finite
width effects on the $\rho$-meson by the pion due to collisional
broadening. It has been extended to the case of relativistic
quantum field theory by Cornwall, Jackiw, and Tomboulis \cite{Cornwall:1974vz}.

This functional method is based on the resummation of the
partition sum. In this scheme one obtains
Dyson Schwinger equations for the full propagators of the
$\rho$-meson and the pions with the following form:

\begin{subequations}
\bea \label{schwingerPiRho} {\cal S}_{\mu \nu}^{-1}(K) &=&
S_{\mu \nu}^{-1}(K) +   \Pi_{\mu \nu}(K)\,\, , \\
{\cal P}_{}^{-1}(K) &=& P_{}^{-1}(K) + \Pi_{}(K)\,\, , \eea
\end{subequations}
where $\cal{S}$ and $\cal{P}$ represent the full propagators of
the $\rho$-meson and the pions, respectively. $S$ and $P$ are the
free propagators of the $\rho$-meson and the pions:
\begin{subequations} \label{treePiRho}
\bea  \label{Drho} S^{-1}_{\mu \nu}(K) &=&
(K^2-m^2)g_{\mu \nu}+\frac{1-\xi}{\xi} K_\mu K_\nu\,\, , \\
P^{-1}(K) &=& K^2-m_{\pi}^2 \,\,. \label{Dpi2} \eea
\end{subequations}

The self-energies $\Pi^{\mu \nu}$ and $\Pi$ are the self-energies
of the $\rho$-meson and the pions, respectively, which are given as
functional derivatives of the sum of the two-particle irreducible
diagrams $V_2$ of the theory in this resummation formalism:

\begin{subequations} \label{selfenergyPiRho}
\bea \Pi(P) &\equiv& \left. 2 \, \frac{\delta V_2 [
        \bar{S}, \bar{P}]}{\delta \bar{P}_{}(P)}
\right|_{\bar{S}={\cal S},\bar{P}={\cal P}} \,\, {\rm and}
\\\Pi_{\mu \nu}(P) &\equiv& \left. 2 \, \frac{\delta V_2 [
\bar{S}, \bar{P}]}{\delta \bar{S}^{\mu \nu}(P)}
\right|_{\bar{S}={\cal S},\bar{P}={\cal P}} \,\, .  \eea
\end{subequations}

\begin{figure}
\begin{center}
\includegraphics[width=8cm]{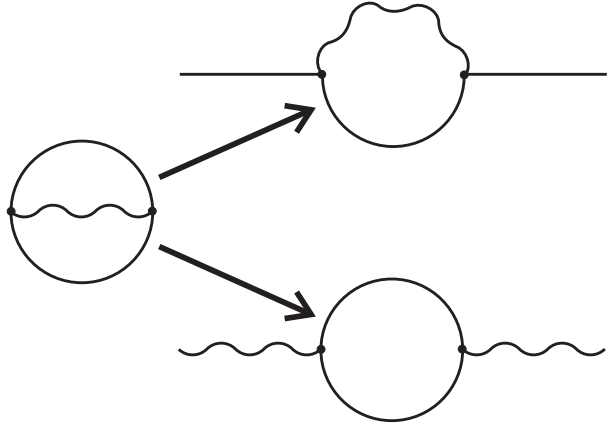}
\caption{On the left the diagram representing the approximation for $V_2$ is
 depicted. Via functional variation one obtains the self-energy contributions
 to the pion and $\rho$-meson self-energies as on the ride hand side. 
 The wiggled line represent the $\rho$-meson propagator and the solid
 line the pion propagator. The diagrams have to be resummed self-consistently.}
\label{diagram}
\end{center}
\end{figure}

Taking all two-particle irreducible diagrams into account
corresponds to solving the full quantum field theory. A systematic
approximation scheme is obtained by taking into account certain
classes of two-particle irreducible diagrams. In our analysis we
consider the following approximation of $V_2$:\bea \label{V2PiRho}
V_{2}=-g^2\int_L \int_P (2P+L)^\mu(2P+L)^\nu \bar{S}_{\mu \nu}(L)
\bar{P}(P) \bar{P}(L+P) \,\,, \eea which corresponds to the
diagram on the left of fig. \ref{diagram}.
Our notation is $
\int_k \, f(k) \equiv T \sum_{n=-\infty}^{\infty}
                        \int d^{3}{\bf k}/(2\pi)^3 \,
         f(2 \pi i n T,{\bf k}) $.
This scheme does not include any resummation
corrections of the vertices, so in $V_2$ one takes the vertices of
the $\pi-\rho$  interaction stemming from the Lagrangian density
(\ref{effLagPiRho2}) on the tree-level.

Calculating the self-energies for example for the $\rho$-mesons
according to eqn. (\ref{selfenergyPiRho}) corresponds on the
diagrammatic level to cutting the $\rho$-propagator in the diagram
of $V_2$. In an analogous way one obtains the self-energy of the
pion  (see fig. \ref{diagram}) . The scheme is self-consistent. Because
every propagator appearing in the self-energy contributions is a full-propagator, i.e. in a perturbative language
the propagators in the loop diagrams are fully resummed on that
loop level.

The self-energies of the pions and the $\rho^0$-meson are
therefore given by (\ref{selfenergyPiRho}):
 \bea \label{selfPion} \Pi(P)=2\frac{\delta
V_2}{\delta {\cal P}(P)}= -4g^2\int_Q (2P-Q)^\mu (2P-Q)^\nu {\cal
S}_{\mu \nu}(Q){\cal P}(P-Q)\nn \\ \eea
 and
\bea \Pi_{\mu \nu}(P)=2\frac{\delta V_2}{\delta {\cal S}^{\mu
\nu}(P)}= -2g^2\int_Q(2Q-P)_{\mu}(2Q-P)_{\nu} {\cal P}(Q) {\cal
P}(P-Q) \,\, .\eea It is obvious that these
self-energies themselves functionally depend on the propagators
which enter the Dyson-Schwinger equations.

As we are interested in the dilepton emission from a  medium in thermal equilibrium, we need to calculate the retarded propagator of the $\rho$-meson in thermal field theory, see e.g. \cite{LeBellac}. That quantity directly enters the calculation of the dilepton production rates.
For that purpose we recall that the thermal average $\langle \hat{A} \rangle$ 
of an operator $\hat{A}$ is defined by
\bea
\langle \hat{A} \rangle_\beta=\frac{1}{Z(\beta)} {\rm Tr}\big( \hat{A}  {\rm e} ^{-\beta \hat{H}}\big)\,\, , 
\eea
where $\hat{H}$ is the Hamiltonian, $\beta=1/T$ is the inverse temperature and $Z(\beta)$ is  the partition sum.
 The retarded propagators of the
$\rho$-meson and the pions are defined as follows:

\bea 
{\cal S}^{\mu
\nu}_{\rm R}(t)&=&-i\langle
\theta(t)[\hat{A}^\mu(t),\hat{A}^\nu(0)]\rangle_{\beta}
\,\, ,  \nn \\
{\cal P}_{\rm R}(t)&=&-i\langle
\theta(t)[\hat{\phi}(t),\hat{\phi}(0)]\rangle_{\beta}\,\,. \eea
They can be expressed in terms of the spectral functions of the
$\rho$-meson and the pions: 

\bea 
{\cal S}^{\mu \nu}_{\rm R}
(k_0,{\bf k})
 &=& 
\int_{-\infty}^{\infty} \frac{dp_0'}{2\pi} 
\frac{\rho^{\mu\nu }(p_0',{\bf k})}{k_0-p_0'+i\eta'} \, \,, 
\nn\\
 {\cal P}_{\rm R}(k_0,{\bf k})&=&
\int_{-\infty}^{\infty} \frac{dp_0'}{2\pi} \frac{\rho(p_0',{\bf
k})}{k_0-p_0'+i\eta'} \, \,,  
\eea
where an analytical continuation from Matsubara frequencies to real energies has
been performed.

The vector meson propagator is a Lorentz-tensor. Lorentz
invariance is broken because the 4-vector of the 
medium defines a preferred
frame of reference and the tensorial
densities (as spectral functions, self-energies, propagators)
become functions of energy and of momentum. To identify the
physically propagating modes of the vector meson, we project its
propagators, self-energies and spectral function onto the
four-dimensionally transverse and longitudinal subspaces. A
detailed overview of the conventions of the projection formalism
is given in appendix \ref{projector} .

Every tensorial quantity can be decomposed into the
four-dimensionally transverse components (proportional to the
projectors A and B) and longitudinal components (proportional to
the projectors C, E), see eqn. (\ref{tensor}).

In general the spectral function components can be expressed
via:\bea \label{spektraldichten} \rho_{\rm A}(k_0,{\bf k})&=& -2
{\rm Im}\,  \frac{1}{2} {\rm A}_{\mu \nu}(k_0,{\bf k})
 D^{{\rm R}\mu \nu}(k_0,{\bf k})= -\,{\rm Im}
\int_{-\infty}^{\infty} \frac{dk_0'}{2\pi} \frac{{\rm A}_{\mu
\nu}(k_0,{\bf k}) \rho^{\mu \nu}(k_0',{\bf k})}{k_0-k_0'+i\eta}
\nn \\ &=& \frac{1}{2} {\rm A}_{\mu \nu} \rho^{\mu \nu}\,\, , \nn \\
\rho_{\rm B}(k_0,{\bf k})&=&  -2 {\rm Im} \,{\rm B}_{\mu
\nu}(k_0,{\bf k})
 D^{{\rm R}\mu \nu}(k_0,{\bf k})= -2 {\rm Im}
\int_{-\infty}^{\infty} \frac{dk_0'}{2\pi} \frac{{\rm B}_{\mu
\nu}(k_0,{\bf k}) \rho^{\mu \nu}(k_0',{\bf k})}{k_0-k_0'+i\eta}
\nn \\ &=&  {\rm B}_{\mu \nu}(k_0,{\bf k}) \rho^{\mu \nu}(k_0,{\bf
k}) \, \, ,
\nn \\
\rho_C(k_0,{\bf k})&=&  -2 \,{\rm Im} \, \left(-\frac{1}{2 k_0^2
k^2}{\rm C}_{\mu \nu}(k_0,{\bf k})
 D^{{\rm R}\mu \nu}(k_0,{\bf k}) \right) \nn \\  &=& \frac{1}{k_0^2 k^2}\,{\rm Im}
\int_{-\infty}^{\infty} \frac{dk_0'}{2\pi} \frac{{\rm C}_{\mu
\nu}(k_0,{\bf k}) \rho^{\mu \nu}(k_0',{\bf k})}{k_0-k_0'+i\eta}
\nn \\ &=&  -\frac{1}{2 k_0^2 k^2} {\rm C}_{\mu \nu}(k_0,{\bf k})
\rho^{\mu \nu}(k_0,{\bf k})
\nn  \, \, ,\\
\rho_E(k_0,{\bf k})&=&  -2\, {\rm Im} \,{\rm E}_{\mu \nu}(k_0,{\bf
k})
 D^{{\rm R}\mu \nu}(k_0,{\bf k})= -2 \,{\rm Im}
\int_{-\infty}^{\infty} \frac{dk_0'}{2\pi} \frac{{\rm E}_{\mu
\nu}(k_0,{\bf k}) \rho^{\mu \nu}(k_0',{\bf k})}{k_0-k_0'+i\eta}
\nn \\ &=&  {\rm E}_{\mu \nu}(k_0,{\bf k}) \rho^{\mu \nu}(k_0,{\bf
k}) \, \, . \eea

The retarded propagator can be rewritten in terms of these
spectral function's components: \bea \label{Dmunu} {\cal S}_{{\rm
R}\mu \nu}(k_0,{\bf k})= \int_{-\infty}^{\infty}
\frac{dk_0'}{2\pi} \frac{ \left(\rho_{\rm A} {\rm A}_{\mu \nu} +
\rho_{\rm B} {\rm B}_{\mu \nu }+ \rho_{\rm C} {\rm C}_{\mu \nu} +
\rho_{\rm E} {\rm E}_{\mu \nu})(k_0', {\bf k}
\right)}{k_0-k_0'+i\eta}. \eea

Inverting the resummed propagator (see eqn. \ref{schwingerPiRho})
of the $\rho$-meson and employing eqns. (\ref{spektraldichten}) leads
to: \bea \label{spektral1} \rho_{\rm A}&=& -2 {\rm Im}\,
\frac{1}{2}{\rm A}_{\mu \nu}(p_0,{\bf p})
 {\cal S}^{\mu \nu}(p_0,{\bf p})=-2\, {\rm Im} \,
 \left[\frac{1}{P^2-m^2+\Pi_a}\right]
 \, \, ,
\nn \\
\rho_{\rm B}&=&  -2 {\rm Im} \,{\rm B}_{\mu \nu}(p_0,{\bf p})
 {\cal S}^{\mu \nu}(p_0,{\bf p})
 = -2 {\rm Im} \, \left[b^{-1}(P)\right]\, \, ,
\nn \\
\rho_{\rm C}&=&  -2 \,{\rm Im} \, \frac{1}{-2 p_0^2 p^2}{\rm
C}_{\mu \nu}(p_0,{\bf p})
 {\cal S}^{\mu \nu}(p_0,{\bf p})\, \, \nn \\
&=& -2 \xi {\rm Im} \,\left[- b(P) \frac{P^2-\xi m^2 +\xi \Pi_e
}{\Pi_c}\right]^{-1}\, \, ,
\nn \\
\rho_{\rm E} &=&  -2\, {\rm Im} \,{\rm E}_{\mu \nu}(p_0,{\bf p})
 {\cal S}^{\mu \nu}(p_0,{\bf p}) \nn \\
 &=&-2 \xi {\rm Im}\,\left[b(P)\frac{P^2-\xi m^2 + \xi \Pi_e}{P^2-m^2+\Pi_b}\right]^{-1}\, \, ,
 \eea where the function $b(P)$ is given by
 \bea
   b(P)=P^2-m^2+\Pi_b+\xi \frac{p_0^2 p^2 \Pi_c^2}{P^2-\xi m^2 +\xi
   \Pi_e}\, \, 
 \eea

and the projected components $\Pi_a$, $\Pi_b$, $\Pi_c$, and
$\Pi_e$ (see appendix \ref{projector}) of the self-energy $\Pi_{\mu \nu}$ are
introduced. The quantity $\xi$ is the gauge parameter appearing in
the Stückelberg method in the $R_\xi$-gauge (see eqn.
(\ref{gaugefix})).

In spite of the truncation of the resummation (e.g. limiting the
sum of the two-particle irreducible diagrams $V_2$ to a subset of
all possible digagrams) conservation of conserved currents
stemming from global symmetries as long as they are realised as
linear representations of the conserved group detailed balance
and unitarity as well as thermodynamical consistence. This is
guaranteed in $\Phi$-functional approximation schemes, as shown by
Baym \cite{Baym:1962sx}. However, in these schemes the vertex corrections
necessary to guarantee Ward Takahashi identies for the propagator
are not taken into account consistently. This shortcoming can lead
to serious problems of the self-consistent treatment of vector
particles on the propagator level \cite{HendrikDiss}.

This problem has to be addressed in our scheme as well. In \cite{vanHees:2003dk} 
it has
been shown that in order to be consistent with the corresponding
Ward Takahashi identities the following relation: \footnote{This
relation is somewhat similar to those for QED, e.g.$ P^\mu {\cal
S}^{\rm QED}_{\mu \nu}=  \xi P_{\nu}/P^2 \, \, $ and QCD, e.g.
$P^{\mu} P^{\nu} {\cal S}^{QCD}_{\mu \nu} =\xi $.}: \bea P^\mu
{\cal S}^{\rm Stueck}_{\mu \nu}  &=& \xi \frac{P_{\nu}}{P^2+\xi
m^2} \, \, \eea for the (fully
resummed) propagator of the massive vector meson $\cal{S}_{\mu
\nu}$ has to be fulfilled in the Stückelberg theory. From that condition one can conclude
that in Stückelberg theories  (as in QED) the self-energy tensor of
the vector particle has to be four-dimensionally transverse which
corresponds to vanishing $\Pi_c$ and $\Pi_e$ components of
the self-energy tensor.

This relation on the correlator level is spoiled in the employed
$\Phi$-functional approximation, since only a partial resummation
is performed. This fact is well known, not only for
Stückelberg theories \cite{vanHees:2000bp}, 
but also from Dyson-Schwinger approaches
to the description of QCD phenomena \cite{vonSmekal:1997is} and stems from a
violation of the gauge symmetry.

One may think that these fundamental considerations of the
violation of Ward Takahashi identies don't matter from a
phenomenological point of view, because we are
interested only in an apropriate description of the $\rho$-meson studying
it's importance for the understanding of dilepton spectra and not
about intrinsically complicated considerations on the violation of
local symmetries in gauge theories. But we calculate 
the production of dileptons in a VMD approach
where the $\rho$-meson couples directly to the virtual photon, these
questions have to be addressed in detail. In part \ref{production} of the
appendix the equation for the rate of dilepton production is
given. Eqn. (\ref{currentRhoProp}) shows that the
current-current correlator is directly proportional to the
retarded propagator of the $\rho$-meson. Current conservation in QED
which is essential for an appropriate description of the dilepton
production tells that the spectral-function of the $\rho$-meson has
to be four-dimensionally transverse: $P_\mu \rho^{\mu \nu}=0$.

To guarantee conservation of
the hadronic electromagnetic current, we work in an appropriate
gauge: the $R_\xi$-gauge taking $\xi \rightarrow 0$
\footnote{That is similar to taking the landau gauge in
Dyson-Schwinger approaches to QCD at finite temperature, see e.g.
\cite{Maas:2002if}.}. This leads to the following modification of the eqns.:
 \bea
\label{spektral2} \rho_{\rm A}&=& -2 {\rm Im}\, \frac{1}{2}{\rm
A}_{\mu \nu}(p_0,{\bf p})
 {\cal S}^{\mu \nu}(p_0,{\bf p})=-2\, {\rm Im} \, \left(
 P^2-m^2+\Pi_a\right)^{-1}
 \, \, ,
\nn \\
\rho_{\rm B}&=&  -2 {\rm Im} \,{\rm B}_{\mu \nu}(p_0,{\bf p})
 {\cal S}^{\mu \nu}(p_0,{\bf p})
 = -2\, {\rm Im} \, \left( P^2-m^2+\Pi_b \right)^{-1}\, \,,
\nn \\
\rho_{\rm C}&=&  -2 \,{\rm Im} \, \frac{1}{-2 p_0^2 p^2}{\rm
C}_{\mu \nu}(p_0,{\bf p})
 {\cal S}^{\mu \nu}(p_0,{\bf p})
= 0 \, \,,
\nn \\
\rho_{\rm E} &=&  -2\, {\rm Im} \,{\rm E}_{\mu \nu}(p_0,{\bf p})
 {\cal S}^{\mu \nu}(p_0,{\bf p}) =0 \, \,.
 \eea The components $\rho_C$ and $\rho_E$ do vanish and the
transversality of the spectral-function $P^\mu \rho_{\mu \nu}=0$
and therefore current conservation is guaranteed. In that way we
solve the most serious problem, mainly former violation of
electromagnetic current conservation. On the level of the
$\rho$-meson self-energy we still have non-vanishing $\Pi_c$ and
$\Pi_e$ components, but this does not lead to problems
calculating the dilepton rates. 
In the following we refer to this technique to guarantee the 
transversality of the spectral functions as "{\it approach I}". 

In \cite{vanHees:2000bp} 
the authors use a different ansatz by forcing 
the self-energy of the $\rho$-meson to be explicitly transverse via the eqns. (\ref{transTensor}) (with $X\equiv\Pi$), which
are violated in $\Phi$-functional approximation schemes. 
In this ansatz 
one argues that the temporal components of the self-energy tensor are tied to the conseration of charge and therefore involve an infinite relaxation time of the conserved quantity in the full theory. The self-consistent results are different because they reflect the damping time of the propagators in the loop. This behaviour has been studied in detail in \cite{Knoll:1995nz}, where it has been shown that current conservation could be restored via a Bethe-Salpeter ladder resummation. The spatial components of the self-energy tensor are in general be less affected by the resummation in the case that the relaxation time for the transverse current-current correlator is comparable to the damping time of the propagators appearing in the loop, whereas the time components are expected to suffer significant changes \cite{vanHees:2000bp}.
This physical reasoning leads to enforcing the transversality of the self-energy tensor via the first two eqns. of (\ref{transTensor}) (with $X\equiv\Pi$), which give the $\rho$-mesons four-dimensionally transverse self-energy components ($\Pi_a$, $\Pi_b$) as functions of $
\Pi^t(P) = \frac{1}{2} \, \left( \delta^{ij} - \hat{p}^i\,
\hat{p}^j \right) \, \Pi^{ij}(P) $ and $ \Pi^\ell(P)
= \hat{p}_i \, \Pi^{ij}(P)\, \hat{p}_j $:
\begin{subequations} 
\begin{eqnarray}
\Pi^{\rm a}(P) & = &  -  \Pi^t(P) \,\, ,  \\
\Pi^{\rm b}(P) & = &  -  \frac{P^2}{p_0^2} \Pi^{\ell}(P) \,\, .
\end{eqnarray}
\end{subequations}
These equations are equivalent to eqn. (4) in  \cite{vanHees:2000bp} up to a different sign convention.  
In the following we refer to this technique  to guarantee the transversality of the self-energy tensor first employed in \cite{vanHees:2000bp} as 
"{\it approach II}".

For the sake of simplicity we are concentrating on damping width effects of the particles and avoid renormalization which would require a temperature
independent self-consistent subtraction of infinities \cite{vanHees:2001ik} 
and don't
take into account any changes in the real part of the self-energies. This ansatz to disregards any mass shift of the particles is analogous to the one in  \cite{vanHees:2000bp}.

Given the fact that we are interested in the phenomenology of the broadening due to finite width of the pions this approximation of neglecting mass shifts (i. e. disregarding  renormalization), can be justified. In \cite{Gale:1991pn} the $\pi$-$\rho$ interaction has been studied perturbatively. The value of the mass shifts (relative to vacuum) predicted from this perturbative study are comparable with the mass shifts one can expect in any many body scheme. (This is different regarding the collisional broadening, since in the perturbative calculation the width of the pion giving raise to the $\rho$ meson's width is zero.)
Especially from Fig. 6, 7, 8 and, Fig. 9 in \cite{Gale:1991pn} one expects physical mass shifts  from the $\pi$-$\rho$-interaction only in the range of about $10\%$ of the vacuum mass of the $\rho$-meson even for comparably high temperatures. 

In that way, the equations for the non-vanishing components of the
spectral function are:
\bea \label{spektral3} \rho_{\rm A}&=&
\frac{-2\,{\rm Im} \, \Pi_a }{(P^2-m^2)^2+({\rm Im} \, \Pi_a)^2}
 \, \, ,
\nn \\
\rho_{\rm B}&=&\frac{-2\,{\rm Im} \, \Pi_b }{(P^2-m^2)^2+({\rm Im}
\, \Pi_b)^2 } \, \,
\nn \\
\rho_{\rm C}&=& 0 \,\,
\nn \\
\rho_{\rm E}&=& 0 \,\,
\eea

To close this self-consistent set of equations, we have to express
the self-energies as functionals of the spectral function of the
pion and the spectral function's components $\rho_a$ and $\rho_b$
of the $\rho$-meson. We use the decomposition of the spectral function of the $\rho$-meson:
\bea
 \rho_{\mu \nu}(P)&=&A_{\mu \nu} \rho_{a}(P) + B_{\mu \nu} \rho_{
 b}(P)= \nn \\
 &=& \bigg(-\delta_{ij}+\hat{p}_i\hat{p}_j\bigg)\rho_{ a}(P)+
 \bigg(g_{\mu \nu} + \delta_{ij}-\frac{P_\mu
 P_\nu}{P^2}-\hat{p}_i\hat{p}_j\bigg) \rho_{\rm b}(P)\, \,, \nn
  \eea
 where $\hat{p}_i$ is the normalized i.th component of the
momentum vector $\hat{p}_i=p_i/p$ and the following contractions
\bea \label{contract2} g_{\mu
\nu}(2P-Q)^{\mu}(2P-Q)^{\nu}&=&2(P^2+L_2^2)-L_1^2 \,\,, \nn \\
\frac{Q_\mu Q_\nu}{Q^2}
(2P-Q)^{\mu}(2P-Q)^{\nu}&=&\frac{(P^2-L_2^2)^2}{L_1^2} \, \, , \nn
\\
 \delta_{ij}(2p-q)^{i}(2p-q)^{j}&=&2(p^2+l_2^2)-l_1^2 \,\,, \nn \\
\frac{q_i q_j}{q^2}
(2p-q)^{i}(2p-q)^{j}&=&\frac{(p^2-l_2^2)^2}{l_1^2} \, \, .
 \eea where $L_1=Q$ and $L_2=P-Q$. The quantities
$l_1^2, l_2^2$, and $p^2$ are the squared three-momenta.
$l_{10}$ and $l_{20}$ are the energy components of $L_1$ and $L_2$, respectively. We 
obtain the following equation of  part of the
self-energy of the pion:\bea \label{ImselfenergyPion} {\rm Im}
\Pi(P) =&&- \pi \frac{4g^2}{p(2 \pi)^2}\int_{0}^{\infty}  l_1 \d
l_1 \int_{0}^{\infty}  l_2 \d l_2 \int_{-\infty}^{\infty}\frac{\d
l_{10}}{2\pi}
\int_{-\infty}^{\infty}\frac{\d l_{20}}{2\pi}  \nn \\
&& \times
\bigg\{-\left[2(p^2+l_2^2)-l_1^2-\frac{(p^2-l_2^2)^2}{l_1^2}\right]\rho_a(l_{10},l_1)\nn\\
&& ~-\big[2(P^2+L_2^2)-L_1^2+2(p^2+l_2^2)-l_1^2
-\frac{(P^2-L_2^2)^2}{L_1^2} \nn \\
&&~ -\frac{(p^2-l_2^2)^2}{l_1^2}\big]\rho_b(l_{10},l_1)\bigg\} \rho(l_{20}, l_2) \nn \\
&&~\times \Theta\left(|l_1-l_2|\leq p\leq l_1+l_2\right) \nn
\\&&~ \times \left[1+f(l_{10})+f(l_{20})\right]
\delta(p_0-l_{10}-l_{20}) \, \, . \eea
 To determine the imaginary
four-dimensionally transverse components of the self-energy
${\rm Im}\, \Pi_a$ and ${\rm Im} \, \Pi_b$, we employ eqns.
(\ref{XProj})(with $X\equiv\Pi$) and use the following
contactions:\bea \label{contract1}
\delta_{ij}(2q-p)^{i}(2q-p)^{j}&=&2(l_1^2+l_2^2)-p^2 \,\,,\nn \\
\frac{p_i p_j}{p^2}
(2q-p)^{i}(2q-p)^{j}&=&\frac{(l_1^2-l_2^2)^2}{p^2}\,\,, \nn \\
(2q-p)^{0}(2q-p)^{0}&=&(l_{10}-l_{20})^2\,\,, \nn \\
(2q-p)^{0}(2q-p)_{i}\frac{p^i}{p}&=& -
\frac{(l_{10}-l_{20})(l_1^2-l_2^2)}{p}\,\,,\eea
We need to calculate ${\rm Im} \left(\delta^{ij}\Pi_{ij}\right)$,
${\rm Im}\left( \hat{p}^i \hat{p}^j \Pi^{ij}\right)$, ${\rm Im}
\left(\Pi^{00}\right)$ and ${\rm Im}
\left(\Pi^{i0}\hat{p}_i\right)$. These are given by:

\begin{subequations}\label{ImselfenergyProjRho}
\bea {\rm Im}\, \left(\delta^{i j}\Pi_{i j}\right)&=&-\pi
\frac{2g^2}{p(2 \pi)^2}\int_{0}^{\infty} l_1 \d l_{1}
\int_{0}^{\infty} l_2 \d l_{2} \int_{-\infty}^{\infty}\frac{\d
l_{10}}{2\pi}
\int_{-\infty}^{\infty}\frac{\d l_{20}}{2\pi}  \nn \\
&&~\times \left[2(l_1^2+l_2^2)-p^2 \right]
\rho(l_{10},l_1)\rho(l_{20},l_2) \delta\left(p_0-l_{10}-l_{20}\right) \nn\\
&&~\times~ \Theta\left(|l_1-l_2|\leq p\leq l_1+l_2\right) \nn \\
&&~\times~\left[1+f(l_{10})+f(l_{20})\right] ,  \\
{\rm Im}\, \left( \hat{p}_{i}\hat{p}_{j}\Pi^{i j}\right)&=&-\pi
\frac{2g^2}{p(2 \pi)^2}\int_{0}^{\infty} l_1 \d l_{1}
\int_{0}^{\infty} l_2 \d l_{2} \int_{-\infty}^{\infty}\frac{\d
l_{10}}{2\pi}
\int_{-\infty}^{\infty}\frac{\d l_{20}}{2\pi}  \nn \\
&&~\times \left[\frac{(l_1^2-l_2^2)^2}{p^2}\right]
\rho(l_{10},l_1)\rho(l_{20},l_2) \delta\left(p_0-l_{10}-l_{20}\right)\nn \\
&&~\times~ \Theta\left(|l_1-l_2|\leq p\leq l_1+l_2\right) \nn \\
&&~\times~\left[1+f(l_{10})+f(l_{20})\right] ,  \\
{\rm Im} \left(\Pi^{00}\right) &=&- \pi \frac{2g^2}{p(2
\pi)^2}\int_{0}^{\infty} l_1 \d l_{1} \int_{0}^{\infty} l_2 \d
l_{2} \int_{-\infty}^{\infty}\frac{\d l_{10}}{2\pi}
\int_{-\infty}^{\infty}\frac{\d l_{20}}{2\pi} \nn \\
&&~\times \left[(l_{10}-l_{20})^2\right] \rho(l_{10},l_1)\rho(l_{20},l_2)\delta\left(p_0-l_{10}-l_{20}\right)\nn \\
&&~\times~ \Theta\left[|l_1-l_2|\leq p\leq
l_1+l_2\right] \nn\\
&&~\times~\left[1+f(l_{10})+f(l_{20})\right] \,\,, \eea and \bea
{\rm Im}\left(\Pi^{i0}\hat{p}_i \right) &=&- \pi \frac{2g^2}{p(2
\pi)^2}\int_{0}^{\infty} l_1 \d l_{1} \int_{0}^{\infty} l_2 \d
l_{2} \int_{-\infty}^{\infty}\frac{\d l_{10}}{2\pi}
\int_{-\infty}^{\infty}\frac{\d l_{20}}{2\pi}  \nn \\
&&~\times \left[-\frac{(l_{10}-l_{20})(l_1^2-l_2^2)}{p})\right] \rho(l_{10},l_1)\rho(l_{20},l_2)\nn \\
&&~\times~ \Theta\left(|l_1-l_2|\leq p\leq
l_1+l_2\right)\delta\left(p_0-l_{10}-l_{20}\right)\nn
\\&&~\times~\left[1+f(l_{10})+f(l_{20})\right] \,\,. \eea
\end{subequations}

With this set of equations for the self-energy and
spectral-density components, we have established our
self-consistently resummed approximation scheme.
We solve this set of equations numerically.

\section{Finite width-effects on the $\rho$-meson}\label{width}

In order to gain insight into the physics employing the spectral modifications, we compare the 
result with the resummational approach of the previous section to perturbative one-loop calculations of the 
$\rho$- (cf. \cite{Geddes:1980nd}) and $\pi$ widths. 

As outlined in the last section, we work in the $\Phi$-functional approximation scheme in the $R_\xi$-gauge with $\xi \rightarrow 0$,   which is a choice guaranteeing decoupling of the St\"uckelberg  and Fadeev Popov ghost from the $\rho$ vector meson field and fulfilling the transversality condition of the electromagnetic current (therefore respecting QED gauge invariance). Nonetheless as mentioned before this cannot cure the violation of the Ward Takahashi identities stemming from the violation of local gauge invariance by the $\Phi$-functional approach. That is why we also compare our method to the results obtained by the enforcement of the transversality condition by disregarding the temporal contributions of the tensor (which is spoiling somewhat the $\Phi$-functional approach) which has been suggested and studied by \cite{vanHees:2000bp} in a similar model. We concentrate on the $\rho$-$\pi$ interaction.

In order to gain information to what extend this influences the effects which we are interested in, we discuss the broadening effects in the full $\Phi$-functional approach (referred to as approach I) and in the transversality-enforcing resummation scheme (referred to as approach II). 

The phenomenological most important difference between the resummation approaches  and the perturbative one is, that  
the perturbatively calculated 
self-energies of the $\rho$-meson show a threshold behavior: the self-energies for timelike momenta vanish if $\sqrt{P^2}<2m_{\pi}$, where $P$ is the four-momentum of the meson relative to the rest frame of the medium. This leads to a characteristic threshold in the dilepton-spectra: only lepton-pairs with invariant masses above $2 m_{\pi}$ are emitted. Considering  the self-energy components of the $\rho$-meson in the self-consistent resummation scheme
employed here shows that the finite width of the pions smears out this perturbative threshold behavior:
the whole spectrum  of invariant masses of the dileptons becomes accessible.
Therefore this threshold ought to be considered as an artefact of the 
perturbative calculation.

\subsection{Self-energies and spectral densities}\label{spectral}

\begin{figure}
\begin{center}
\includegraphics[width=17cm]{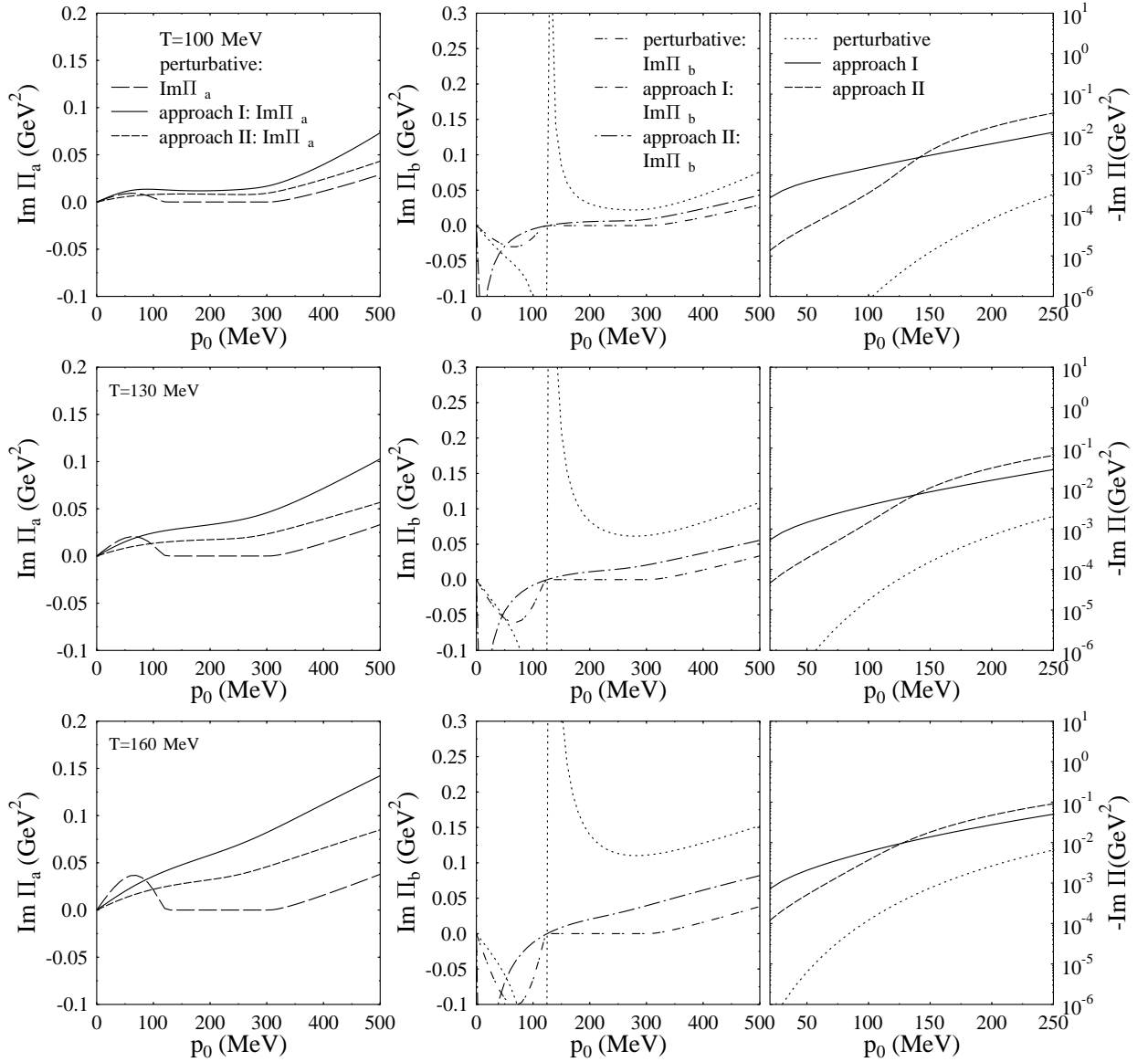}
\caption{The imaginary parts of the spatial-transverse and spatial-
longitudinal self-energies $\Pi_a$ and $\Pi_b$, respectively, as a function
of energy at a constant momentum of $p=125~{\rm MeV}/c$ on the left
hand side and in the middle. We show the results for a perturbative calculation and
for the self-consistent calculations using the full $\Phi$-functional approach (approach I) and
the transversality-enforcing approach (approach II).
The imaginary part of the negative of the pion's self-energy is
depicted on the right hand side. The perturbative 
self-energy of the pion is too small to be seen on the scale of the
graph on the right hand side.}
\label{selfenergy}
\end{center}
\end{figure}

We start our discussion with the four-dimensionally transverse self-energy components ${\rm Im} \Pi_a$ and ${\rm Im} \Pi_b$  of the $\rho$-meson. The ${\rm Im} \Pi_a$-component is spatial-transverse and the ${\rm Im} \Pi_b$-component is spatial-longitudinal. These components are plotted as a function of energy for a fixed momentum $p=125 {\rm MeV/c}$ relative 
to the medium for three different temperature in fig. \ref{selfenergy}(left and middle).

In both self-consistent approaches the spatial transverse and spatial longitudinal imaginary parts of the self-energies show that there is no 
two-pion threshold. The perturbatively calculated imaginary parts 
of the self-energies can be compared to the 
self-consistent ones and vanish for $0<\sqrt{P^2}=\sqrt{p_0^2-p^2}<2 m_{\pi}$. 

The self-energy projectors are (up to possible poles and cuts) 
analytic functions in the complex
energy-plane. Possible cuts on the real energy-axes corresponding to the imaginary part
of the self-energy projections
separate the upper and lower half-planes \cite{LeBellac}. 
Because of the anti-symmetry of the imaginary-parts of the self-energy projectors these cuts extend in the self-consistent
calculations (approach I and approach II) over the total axes. The retarded functions in the upper half plane are therefore
totally separated from the advanced in the lower plane.
This leads in both self-consistent approaches to a (static) 
dilepton-rate that has no two-pion-threshold 
behavior as discussed below.

In approach I the ${\rm Im \Pi}_b$-component shows another remarkable property. 
It has not only a change of sign at
the light-cone (that means 
at the transition between time- and spacelike momenta for $p_0=p$), but
also a singularity. The transversality-condition $P^{\mu}\Pi_{\mu \nu}$ 
means on that level that for $P^2=0$ the component ${\rm Im} \Pi_{b}$ has to vanish. Indeed, that is not fulfilled in the $\Phi$-derivable approach (approach I), but is enforced  in approach II. In approach II the self-energy component ${\rm Im} \Pi_b$ remains a continous function of 
$p_0$ even in the vincinity of the light-cone.
   
Eqn.(\ref{Pib}) shows that the energy component can have a $1/P^2$-pole that leads to a change of sign in ${\rm Im} \Pi_b$ at $p_0=p$ as obsvered in approach I.  
This is an artefact of approach I that is closely correlated to
the violation of the transversality condition of the self-energy tensor $\Pi_{\mu \nu}$, see also the discussion in section \ref{scheme}. Although transversality of the self-energy components  is violated,
the spectral-density tensor $\rho_{\mu \nu}$ still remains four-transverse in the $R_{\xi}$-gauge for $\xi \rightarrow 0$ guaranteeing the conservation of the
hadronic-electromagnetic current. Therefore approach I is consistent within the vector meson dominance approach with quantum electrodynamics, as already discussed above. The singularity of strength $1/P^2$ of the ${\rm Im \Pi}_b$-component at the light-cone still leads to a continuous behavior 
of $\rho_b$ at $p_0=p$.

\begin{figure}
\begin{center}
\includegraphics[width=10cm]{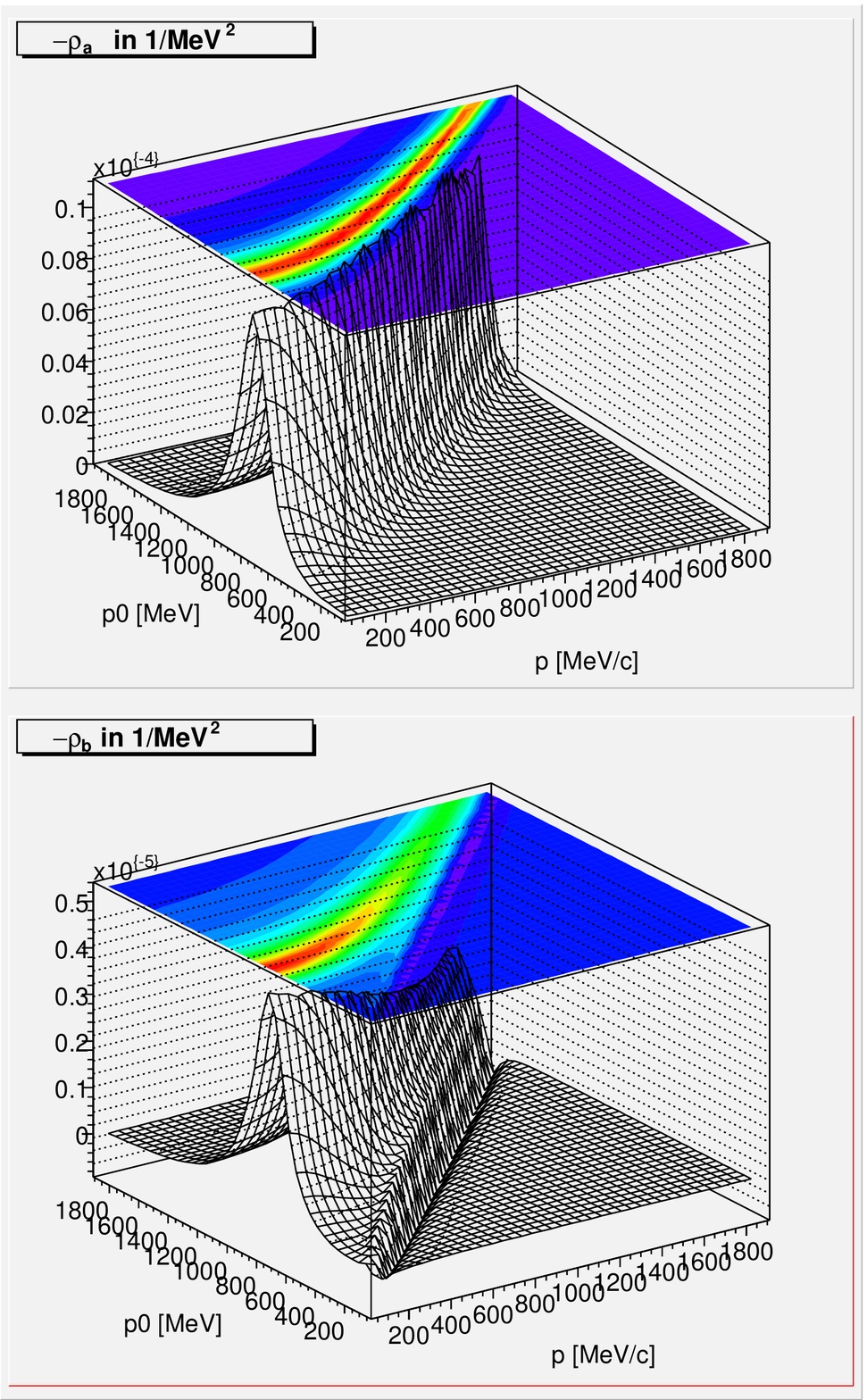}
\caption{(Color online) The spatial-transverse and spatial-
longitudinal spectral functions $\rho_a$ and $\rho_b$, respectively, 
of the rho meson as a function of energy and momentum on at a temperature
of 100 MeV in the full $\Phi$-functional approach (approach I).}
\label{spectralrhoall}
\end{center}
\end{figure}

In figure \ref{spectralrhoall} we plot the components of the spectral-function of the $\rho$-meson 
at a temperature $T=100 \, {\rm MeV}$ as a function of energy $p_0$ and momentum $p$ as obtained within the self-consistent approach I.

\begin{figure}
\begin{center}
\includegraphics[width=15cm]{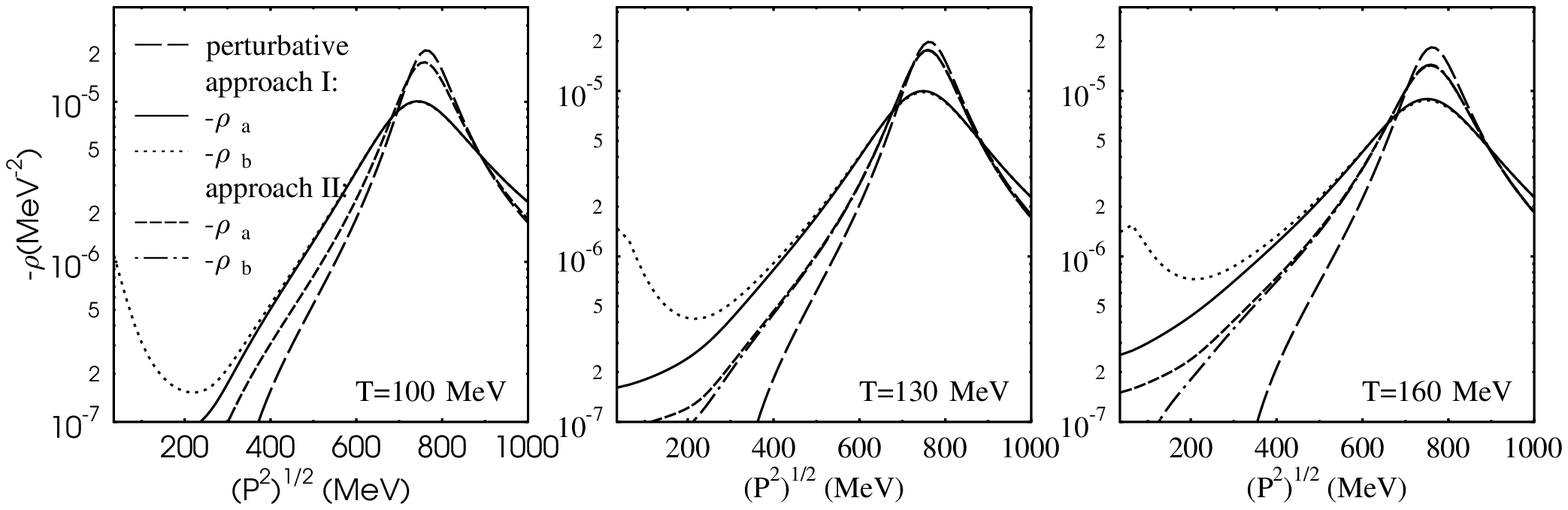}
\caption{The spatial-transverse and spatial-
longitudinal spectral functions $\rho_a$ and $\rho_b$, respectively, 
of the rho meson are given as a function
of invariant mass at a constant momentum of $p=125 MeV/c$ on the left
hand side. We give the results for the perturbative calculation and
for the self-consistent appoaches I and II (see text). The perturbative spatial-transverse and spatial-longitudinal components are within the resolution of the figure identical, that is why we used only one curve for both of them.}
\label{spectralrho}
\end{center}
\end{figure}

We plot the corresponding spatial-transverse and spatial-longitudinal components of the spectral function of the $\rho$-meson depending on $\sqrt{P^2}=\sqrt{p_0^2-p^2}$ for a given momentum of $p=125~{\rm MeV/c}$ for three different temperatures, the two self-consistent approaches and the perturbative calculation in figure \ref{spectralrho}.

\begin{figure}
\begin{center}
\includegraphics[width=10cm]{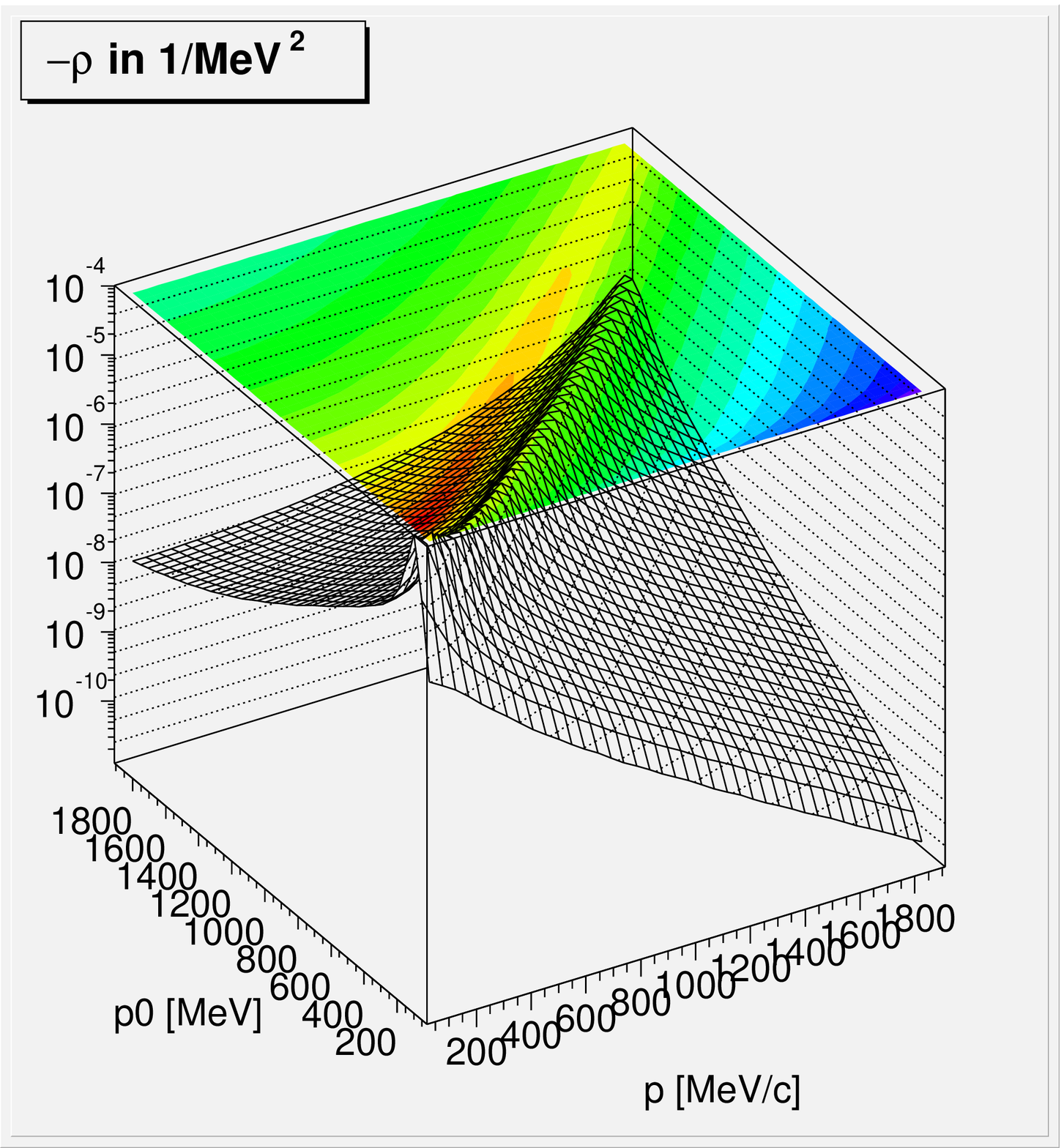}
\caption{(Color online)The spectral function of the pion as 
as a function of energy and momentum at a temperature
of 100 MeV as obtained within the self-consistent calculation applying approach I (see text).}
\label{spectralpionall}
\end{center}
\end{figure}

\begin{figure}
\begin{center}
\includegraphics[width=15cm]{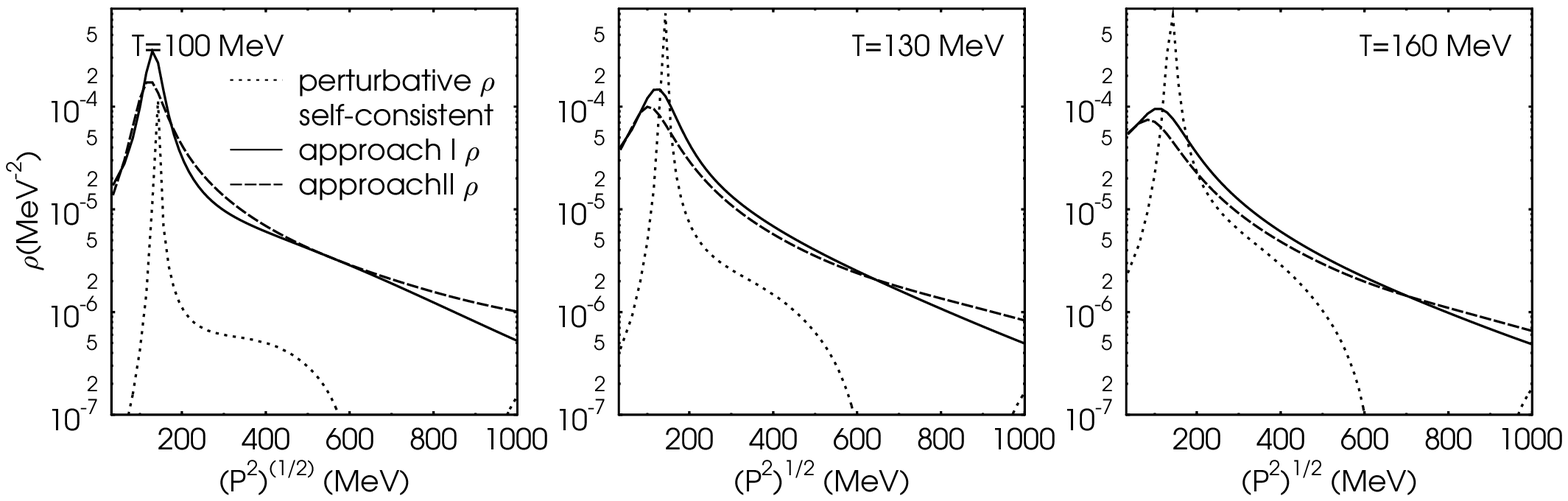}
\caption{The spectral functions $\rho$ 
of the pion is given as a function
of invariant mass at a constant momentum of $p=125 MeV/c$ on the left
hand side. We give the results for the perturbative calculation and
for the self-consistent one.
The width of the pion
right hand side for the self-consistent and the perturbative
calulation as a function of invariant mass.}
\label{spectralpion}
\end{center}
\end{figure}

The spatial longitudinal and transverse components differ in both self-consistent calculations  from each other only in the region about $0<\sqrt{P^2}<500 \, {\rm MeV}$
and are for about $\sqrt{P^2}>500 \, {\rm MeV}$ identical. With rising 
temperature this difference in the lower mass region is enhanced and 
the spectral components get in both self-consistent calculations considerably broadened. The perturbatively calculated spatial transverse and longitudinal spectral components (which are within the resultion of the figure almost identical) are considerably narrower and 
show no spectral strength below the two-pion-threshold. 
In figure \ref{spectralpionall} we show the components of the spectral-function of the pion 
at a temperature of $T=100 {\rm MeV}$ as a function of energy $p_0$ and momentum $p$ for the self-consistent calculation within approach I.
Fig.~\ref{spectralpion} shows the $\pi$ spectral function for different T and a constant 
momentum of $p=125~{\rm MeV/c}$. It shows considerable broadening of the pion in both self-consistent approaches. 
Comparing to the perturbative calculation one oberserves apart from broadening the vanishing of
the spectral function in the range between roughly $630$ and $910$ ${\rm MeV}$.
This is due to the fact that for two particles with mass $m_1>m_2$ the 
perturbative calculation leads to a vanishing width for $m_1-m_2<\sqrt{P^2}<m_1+m_2$ - which corresponds to the two-pion threshold in the case of the 
$\rho$-meson and to the threshold behavior as mentioned here for the pions. 
This threshold behavior is no longer present in both self-consistent approaches.

\subsection{Static thermal dilepton rate} \label{static}

\begin{figure}
\begin{center}
\includegraphics[width=15cm]{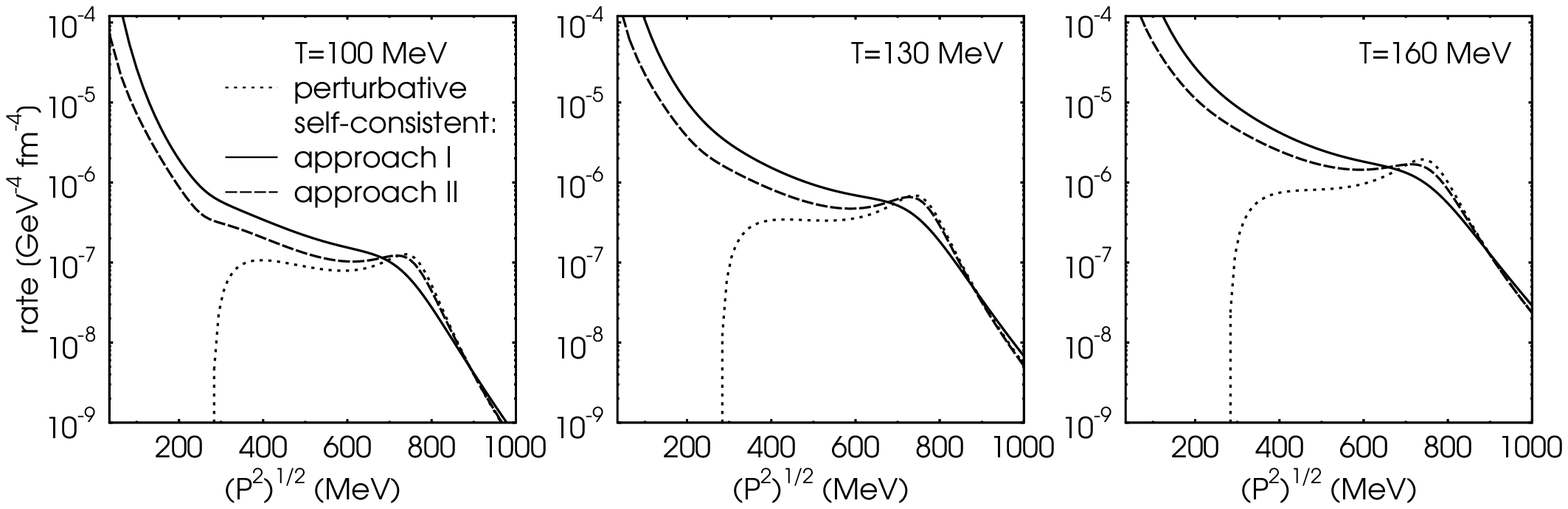}
\caption{The figure shows the static rate of dilepton 
(dielectron)
production as a function of invariant mass. The decaying rho meson
has a constant momentum of $p=125 MeV/c$. We give the self-consistent
results as obtained within approach I and approach II (see text for explanation) and the perturbative ones.}
\label{dileptonstatic}
\end{center}
\end{figure}

We proceed by calculating the static thermal dilepton emission rate in the model. For this purpose, we  
use eqn.(\ref{rateprod}), where this quantity for the hadronic channel is given as a function 
of the invariant dilepton mass and as a functional of the $\rho$-meson 
spectral function based on VMD (see appendix \ref{production} for details). 
We calculate the rate for 
different T  (see figure \ref{dileptonstatic}) for 
decaying $\rho$-mesons of momentum $p=125 {\rm MeV/c}$ relative to
the medium. 

The dilepton emission rate increases with temperature. This enhancement 
is due to the 
Bose Einstein factor in eq.~(\ref{rateprod}). The self-consistently 
calculated rates in approach I and II give rise to a larger yield in the low-mass part
$M=\sqrt{P^2}<800\, {\rm MeV}$ of the spectrum than the perturbatively 
calculated rate which moreover 
has the two-pion threshold below $2 m_{\pi}$.
In the present model, this leads
to a considerable enhancement of the dilepton-rate especially in 
the $\sqrt{P^2}<700\,{\rm MeV}$ sector in comparison 
to the perturbative predictions. The self-consistently calculated 
rates of approach I exhibit even larger broadening effects for the $\rho$-meson than those 
of approach II. Despite this difference both self-consistent approaches show clearly that in this part of the low mass dilepton region the collisional broadening effects enhance the expected static rates considerably and  outweigh the two-pion threshold of the perturbative approach. 
Above $850\, {\rm MeV}$ there
are no important difference in comparison with the perturbative results.

\section{Dilepton emission and fireball evolution}

In order to demonstrate that the results of the previous section, especially the considerable collisional broadening of the $\rho$-meson leads to significant quantitative changes in the prediction of dilepton rates we model the fireball evolution and fold it with the rate calculations.

We emphasize that we study a simplified model of the medium considering only the self-consistent finite width effects of the $\rho$-$\pi$ interaction at finite T where other effects (like mass shifts, chiral symmetry, and scattering off by baryons) have not been included. We do not suggest that these effects are unimportant but wish to demonstrate the self-consistent description of collisional broadening already implies siginificant modifications.

\subsection{The Fireball model}
\label{fireball}

In order to obtain the complete emitted spectrum  we use a mode for the fireball which has been shown to reproduce a number
of other important observables \cite{Renk:2004cj}. Here, we
briefly outline the framework.

The fundamental model assumption is that the fireball matter is in local
thermal (but not necessarily in chemical) equilibrium from an
initial proper-time scale $\tau_0$ till a later breakup at time 
$\tau_f$. We chose volumes corresponding to a
given proper time $\tau$ for the calculation of thermodynamics.
Assuming spatial homogeneity, temperature $T$, entropy density $s$, 
pressure $p$ and chemical
potentials $\mu_i$ as well as energy-density $\epsilon$ become
functions of $\tau$ in such a system.

The volume is assumed to be cylindrically symmetric around the
z-axis of the beam. It is characterized by the longitudinal extension
$L(\tau)$ and the transverse radius $R(\tau)$:
\begin{equation}
\label{E-Volume1} V(\tau) = \pi L(\tau) R^2(\tau).
\end{equation}

In order to account for collective flow effects, we boost
individual volume elements according to a position-dependent
velocity field. For the transverse flow, we make the ansatz
\begin{equation}
\eta_T(r, \tau) = r/R_{rms}(\tau)\eta_T^{rms}(\tau)
\end{equation}
where $R_{rms}(\tau)$ denotes the root mean square radius of the
fireball at $\tau$ and $\eta_T^{rms}(\tau)$ the transverse
rapidity at $R_{rms}$.

For the longitudinal dynamics, we start with the experimentally
measured width of the rapidity interval of observed hadrons
$2\eta_f^{front}$ at breakup. From this, we compute the
longitudinal velocity of the fireball front at kinetic freeze-out
$v_f^{front}$. We do not require the initial expansion velocity
$v_0^{front}$ to coincide with $v_f^{front}$ but instead allow for
a longitudinally accelerated expansion. This implies that during
the evolution $\eta = \eta_s$ is not valid (with $\eta_s$ the
spacetime rapidity $\eta_s = 1/2 \ln ((t+z)/(t-z))$.

The requirement that the acceleration should be a function of
$\tau$ and that the system stays spatially homogeneous for all
$\tau$ determines the velocity field uniquely if the motion of the
front is specified. We solve the resulting system of equations
numerically \cite{Renk:2004cj}. We find that for not too large
rapidities $\eta < 4$ and accelerations volume elements
approximately fall on curves $const. = \sqrt{t^2 - z^2 }$ and that
the flow pattern can be approximated by a linear relationship
between rapidity $\eta$ and  spacetime rapidity $\eta_s$ as
 $\eta(\eta_s) = \zeta\eta_s$
where $\zeta = \eta^{front}/\eta_s^{front}$ and $\eta^{front}$ is
the rapidity of the cylinder front. In this case, the longitudinal
extension can be found calculating the invariant volume $V = \int
d\sigma_\mu u^\mu$ as
\begin{equation}
L(\tau) \approx 2 \tau \frac{\text{sinh }((\zeta -1)
\eta_s^{front}(\tau))}{(\zeta -1)}
\end{equation}
with $\eta_s^{front}(\tau)$ the spacetime rapidity of the cylinder
front. This is an approximate generalization of the
boost-invariant relation $L(\tau) = 2 \eta^{front} \tau$ which can
be derived for non-accelerated motion.

\subsubsection{Parameters of the expansion}

In order to proceed, we have to specify the longitudinal
acceleration $a_z(\tau)$ (which in turn is used to calculate
$\eta_s^{front}(\tau)$ numerically), the initial front velocity
$v_0^{front}$ and the expansion pattern of the radius $R(\tau)$ in
proper time.

For the acceleration  we make the ansatz
\begin{equation}
a_z = c_z \cdot \frac{p(\tau)}{\epsilon(\tau)}
\end{equation}
which allows a soft point in the EoS where the ratio $p/\epsilon$
gets small to influence the acceleration pattern. $c_z$ and
$v_0^{front}$ are model parameters governing the longitudinal
expansion and fit to data.  Hence the EoS translates into a temperature
(and proper-time $\tau$) and determines the acceleration profile $a=const. p(T)/\epsilon(T)$.

Since typically longitudinal expansion is characterized by larger
velocities than transverse expansion, i.e. $v_z^{front} \gg
v_T^{front}$, we treat the radial expansion non-relativistically.
We assume that the radius of the cylinder can be written as
\begin{equation}
R(\tau) = R_0 + c_T \int_{\tau_0}^\tau d \tau'
\int_{\tau_0}^{\tau'} d \tau'' \frac{p(\tau'')}{\epsilon(\tau'')}
\end{equation}

The initial radius $R_0$ is taken from overlap calculations. This
leaves a parameter $c_T$ determining the strength of transverse
acceleration which is also fit to data. The final parameter
characterizing the expansion is its endpoint given by $\tau_f$,
the breakup proper time of the system.

\subsubsection{Evolution of temperature}

We assume that entropy is conserved throughout the thermalized
expansion phase. Therefore, we start by fixing the entropy per
baryon from the number of produced particles per unit rapidity
(see e.g. \cite{Letessier:1993hi}). Calculating the number of
participant baryons (see \cite{Renk:2002md}) we find the total
entropy $S_0$. The entropy density at a given proper time is then
determined by $s=S_0/V(\tau)$.

We describe the EoS in the partonic phase by a quasiparticle
interpretation of lattice data which has been shown to reproduce
lattice results both at vanishing baryochemical potential $\mu_B$
and finite $\mu_B$ \cite{Schneider:2001nf}.

For the phase transition temperature, we choose $T_C = 170$ MeV
based on lattice QCD computations at finite temperature for case
of two light and one heavy quark flavour \cite{Karsch:2000kv}. We also note
that no large latent heat is observed in lattice calculations for the transition and model
the actual thermodynamics as a crossover rather than a sharp phase
transition. Nevertheless, in the calculation we assume quarks and
gluons as degrees of freedom above $T_C$ and hadrons below to
simplify computations. Since the time the system spends in the
vicinity of the transition temperature is small compared with the
total time for dilepton emission, any error we make by
this assumption is bound to be small as soon as we consider the
measured rates which represent an integral over the time evolution
of the system folded with the emission rate.

Since a computation of thermodynamic properties of a strongly
interacting hadron gas close to $T_C$ is a very difficult task, we
follow a simplified approach instead: We calculate
thermodynamic properties of a hadron gas at kinetic decoupling
where interactions cease to be important. Here, we have reason to
expect that an ideal gas will be a good description and obtain
the EoS using an ideal resonance gas model. Using the
framework of statistical hadronization \cite{Renk:2002sz}, we
determine the overpopulation of pion phase space by pions from
decays of heavy resonances created at $T_C$ and include this
contribution (which gives rise to a pion-chemical potential of
order $\mu_\pi \approx 120$ MeV into the calculation). We then
choose a smooth interpolation between decopling temperature $T_f$
and transition temperature $T_C$ to the EoS obtained in the
quasiparticle description. This is described in greater detail in
\cite{Renk:2002md}.

With the help of the EoS and $s(\tau)$, we are now in a position
to compute the parameters $p(\tau), \epsilon(\tau), T(\tau)$ as
well. Since the ratio $p(\tau)/\epsilon(\tau)$ appear in the
expansion parametrization, we have to solve the model
self-consistently.

\subsubsection{Solving the fireball-evolution-model}

In order to adjust the model parameters, we compare with data on
transverse momentum spectra and HBT correlation measurements. This
is discussed in greater detail in \cite{Renk:2003gn, Renk:2004cj}.

By requiring $R(\tau_f) = R_f$ and $v_T^{front} = v_{\perp f}$ we
can determine the model parameters $c_T$ and $\tau_f$. $c_z$ is
fixed by the requirement $\eta^{front} (\tau_f) = \eta_f^{front}$.
The remaining parameter $v_0^{front}$ now determines the volume
(and hence temperature) at freeze-out and can be adjusted such
that $T(\tau_f) = T_f$.

The model for 5\% central 158 AGeV Pb-Pb collisions at SPS is
characterized by the following scales: Initial long. expansion
velocity $v_0^{front} = 0.54c$, thermalization time $\tau_0 = 1$
fm/c, initial temperature $T_0 = 300$ MeV, duration of the QGP
phase $\tau_{QGP} = 6.5$ fm/c, duration of the hadronic phase
$\tau_{had} = 8.5$ fm/c, total lifetime $\tau_f - \tau_0 = 15$ fm/c,
r.m.s radius at freeze-out $R_f^{rms} = 8.55$ fm, transverse
expansion velocity $v_{\perp f} = 0.57 c$.

For the discussion of dileptons, we require the fireball evolution
for other than 5\% central collisions. In this case, we make use
of simple scaling arguments based on the initial overlap geometry
and the number of collision participants. For a detailed
description, see \cite{Renk:2002md, Polleri:2003kn}.

In \cite{Renk:2002md}, it has been shown that this scenario is able
to describe the measured spectrum of low mass dileptons, and in
\cite{Renk:2002sz} it has been demonstrated that under the
assumption of statistical hadronization at the phase transition
temperature $T_C$, the measured multiplicities of hadron species
can be reproduced. In \cite{Polleri:2003kn}, the model has been shown to
describe charmonium suppression correctly. None of these
quantities is, however, very sensitive to the detailed choice of
the equilibration time $\tau_0$. Therefore, we have only
considered the `canonical' choice $\tau_0 =$ 1 fm/c so far. The
calculation of photon emission within the present framework
provides the opportunity to test this assumption and to limit the
choice of $\tau_0$. In \cite{Renk:2003fn}, this has been investigated
in some detail. Within the present framework, the limits $0.5$
fm/c $< \tau_0 < $ $3$ fm/c could be found. Variations within
these limits, however, do not affect the spectrum of dileptons
with invariant mass below 1 GeV significantly.

\subsection{Dileptons from the QGP Phase}\label{QGP}
In order to take into account the contributions from the phase of
a QGP, we employ a quasiparticle picture \cite{Schneider:2001nf}. 
The model
treats quarks and gluons as massive thermal quasiparticles. The
QCD dynamics is incorporated in the thermal masses of these
'effective' quarks and gluons. The masses in this model
approach the hard thermal loop (HTL) results at high temperature.
Around $T_c$ a power law fall-off is assumed for the thermal
masses, based on the conjecture that the phase transition is
either weakly first order or second order. The thermodynamic
quantities of the QGP are calculated in terms of two functions
$B(T)$ and $C(T)$ introduced in the model, see ref. \cite{Schneider:2002tk,
Renk:2002md} , which
account for the thermal vacuum energy and a phenomenological
description of deconfinement.
 As long as the thermodynamically active degrees of freedom are quarks and gluons in a quasiparticle gas,
the virtual photon couples electromagnetically only to the
thermally excited $q \bar{q}$-states 
populating the plasma phase
and converts into a lepton pair, for details see section 4.1 of
\cite{Renk:2002md}. 
This mechanism can be compared to the coupling of the
virtual photon to the vector-meson in VMD in the hadronic phase.

As the preferred reference frame of hot matter breaks Lorentz
invariance, new partonic excitations such as longitudinal gluonic
plasmons or helicity-flipped plasminos could be present. Their
spectral strength remains exponentially suppressed for high
temperatures and momenta which dominate macroscopic thermodynamic
quantities (pressure, entropy and energy density) and their
contribution is neglected in a quasiparticle description. However,
for soft-dilepton production these plasmino modes lead to sharp,
distinct structures in the static dilepton-emission spectra, so
called van Hove singularities \cite{Braaten:1990wp} . 
These peaks should be roughly
located at twice the thermal quasiparticle quark mass $\approx 2
m_q(T)$ and would be smeared out by the fireball evolution, since
the thermal quark masses drop with temperature. Furthermore, since in
the quasiparticle model \cite{Schneider:2001nf} $m_q(T)$ is of order $T$ (smeared out)
van Hove singularities could only appear at low invariant masses and would be
overwhelmed by the hadronic part of the dilepton production.
Therefore the effects of neglecting gluonic and (helicity-flipped)
plasmino modes are expected to be neglible.

\subsection{Dilepton Emission}
\label{emission}

Comparison with the measured dilepton spectrum is enabled by folding
the differential rate $dN/(d^4x d^4q)=dR/d^4q$
with the space-time history of the collision. We want to compare the model
predictions with the CERES/NA45 data \cite{Agakishiev:1995xb,Agakishiev:1997au,Lenkeit:1999xu}
taken in Pb-Au collisions at 158 AGeV (corresponding to a c.m.
energy of $\sqrt{s} \sim 17$ AGeV) and 40 AGeV ($\sqrt{s} \sim 8$
AGeV).  The CERES experiment is a fixed-target experiment. In the
lab frame, the CERES detector covers the limited rapidity interval
$\eta = 2.1-2.65$, {\em i.e.} $\Delta\eta = 0.55$. We integrate
the calculated rates over the transverse momentum $p_T$ and
average over $\eta$, given that $d^4p = M p_T \ dM  \ d\eta \ dp_T
\ d\theta.$ The formula for the space-time- and $p$-integrated
dilepton rate in the framework outlined above hence becomes

\begin{equation}
\frac{d^2N}{dM d\eta} =  \frac{2\pi M}{\Delta \eta} \int
\limits_{\tau_0}^{\tau_{f}} d\tau \  \int  d\eta' \
V(\eta',T(\tau))\int \limits_0^\infty dp_T \ p_T
 \ \frac{dN(T(\tau),M, \eta',
p_T)}{d^4 x d^4p} \ Acc(M, \eta', p_T), \label{integratedrates}
\end{equation}

where $\tau_{f}$ is the freeze-out proper time of the collision.
$V(\eta,T(\tau))$ describes the proper time evolution of volume
elements moving at different rapidities and the function $Acc(M,
\eta, p_T)$ accounts for the experimental acceptance cuts specific
to the detector. At the CERES experiment, each electron/positron
track is required to have a transverse momentum $p_T > 0.2$ GeV,
to fall into the rapidity interval $2.1 < \eta < 2.65$ in the lab
frame and to have a pair opening angle $\Theta_{ee} > 35$ mrad.
Finally, for comparison with the CERES data, the resulting rate is
divided by $dN_{ch}/d\eta$, the rapidity density of charged
particles.

\subsection{Hadronic cocktail in the vacuum}
\label{hadronic}

After freeze-out there are still hadrons present which
decay after a finite time in vacuum and thus contribute to the dilepton
yield.
Below approximately $400$ MeV Dalitz decays are the dominant decay mechanism in 
this regime. We use the experimental analysis of the CERES 
collaboration for
SPS conditions. 

Most notably, remaining $\omega$ and $\phi$ mesons
contribute a characteristic sharp peak structure due to their vacuum
decays. We calculate their contribution directly from the model.

The direct decay of a vector meson $V$ into lepton pairs is given via:
$$
\frac{dN_V}{dM d\eta} = \frac{1}{\Delta \eta} \ \frac{\alpha^2}{12\pi^4} \ R_V(M, T=0) 
\int_{\tau_{f}}^\infty d\tau \ V_{f} \int d^3q \ \frac{M}{q^0} \ f_B(q^0, T_{f}) 
\exp\left(- \frac{\tau - \tau_{f}}{\gamma(q)\tau_0^V}\right).
$$

The temperatur $T_{f}=T(\tau_{f})$ and the volume $V_{f}=V(\tau_{f})$ of the the system
 are to be considered at freeze-out time $\tau_{f}$. 
After this time all medium effects vanish and the {\em vacuum} 
spectral function of the (virtual) photon 
$R_V(M, T=0)$ determines the rate of dilepton production.
The momentum distribution of the vector mesons is given by the Bose-Einstein contribution
determined by the temperature at freeze-out.
Due to the decay of the mesons and the absence of thermal recombination effects the number
of vector mesons is decreasing exponentially as
$\exp\left(- (\tau - \tau_{f})/(\gamma(q)\tau_0^V)\right)$. 
Here, $\tau_0^V$ is the vacuum life time of the vector meson. The quantity $\gamma(q)$
accounts for time dilatation effects on particles with finite three-momentum: 
$$
\gamma(q) = \frac{1}{\sqrt{1 - v^2}} = \frac{q^0}{M}.
$$
Integrating over proper time one obtains:
$$
\frac{dN}{dM d\eta} = \frac{1}{\Delta \eta} \frac{\alpha^2}{12\pi^4}  \tau_0 R(M, T=0)  V_{f} \int d^3q \ f_B(q^0, T_{f}).
$$
The averaged four-dimentional space-time volume available after freeze-out is as one would
expect: $V_{f} \tau_0^V$. 
The three-momentum integral yields the particle density of virtual photons as a function
of the invariant mass: $n(M) = N(M)/V$.
Multiplying the factor $V_{f}$ one obtains the total number of virtual photons decaying in the
freeze-out stage.Weighting this with $R_V(M)$ this gives the
$dN/dM$ distribution for the process in the freeze-out stage vector-meson $\rightarrow 
\gamma^* \rightarrow e^+e^-$.

\section{Dilepton invariant mass spectra} \label{massspectrum}

\begin{figure}
\begin{center}
\includegraphics[width=13cm]{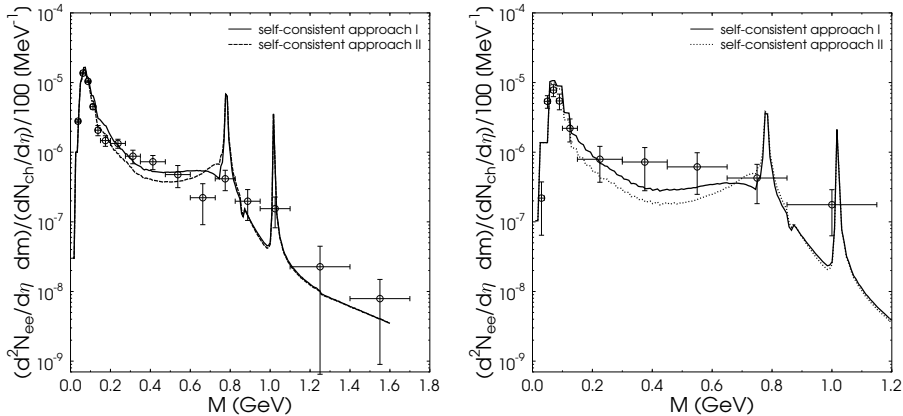}
\caption{Dilepton invariant mass spectra normalized
to $dN_{ch}/d\eta$ in units of $(100 {\rm MeV})^{-1}$ for
the SPS-CERES/NA45 Pb(158 AGeV)+Au (left side) 
and SPS-CERES/NA45 Pb(40 AGeV)+Au (right side) experiment.
Shown are the data, the total rate.
The theoretical spectra are calculated assuming perfect
detector resolution. }
\label{spectra}
\end{center}
\end{figure}

Once the time evolution is given in terms of the temperature $T(\tau)$, baryon
density $\rho(\tau)$ and the knowledge of the dilepton rates from the hadronic
 medium at finite temperature and at freeze-out as well as the rates from
QGP, all necessary ingredients for a quantitative calculation of the dilepton
rates using eqn. 
(\ref{integratedrates}) are in place. 
We consider the SPS CERES/NA45 data taken at beam energies of $40 {\rm AGeV}$ and $158 {\rm AGeV}$. 
The results for the dilepton invariant mass spectra for both experiments 
are shown in fig. \ref{spectra}.

In the calculation we have assumed perfect detector resolution
in invariant mass in order to demonstrate how the $\omega$ and $\phi$ vacuum contributions would
show although these structures in the spectrum cannot be resolved in the experimental
analysis.

Shown in Fig. \ref{spectra} are the results of our model calculations as obtained in the two different self-consistent approaches. The solid line indicates the results in the full $\Phi$-functional approach (approach I), the dashed line the results in the approach where the transversality condition of the self-energy tensor is enforced (approach II).
Both approaches lead to significant contributions for low invariant masses in the 
range of $200 \, {\rm MeV}<M<700 \, {\rm MeV}$. This is due to the collisional broadening of the $\rho$-meson by the $\pi$-$\rho$ interaction in the thermal medium which is taken into account by the self-consistent calculations. This principal finding remains in both self-consistent approaches the same. 

Comparing approach I and approach II one finds  - beyond the confirmation of this in-medium collisional broadening - both results differ quantiatively in the $200 {\rm MeV}<M<800 {\rm MeV}$  region as already expected from the lower static
dilepton rates of approach II in this mass region. These differences result from the two different possible methods to guarantee electromagnetic current conservation in the resummation schemes as discussed above
which can ultimately be traced back to the lack of a feasible resummation scheme which
guarantees local gauge invariance. We emphasize that this ambiguity in the results does not alter the principle finding, namely the importance of self-consistent collisional broadening effects induced by the $\pi$-$\rho$ interaction for an appropriate understanding of the observed spectra.

In order to demonstrate that these broadening effects are quantitatively of the order of magnitude that is important for an experimental understanding, we compare them
 the results to the $40 \, {\rm AGeV}$ and $ 158 \, {\rm AGeV}$ data. We emphasize that many caveats have to be kept in mind comparing these results to data: we consider only the effects due to collisional broadening of the pions and the $\rho$ mesons via their interaction in a simplified model and have not taken into account effects like scattering off by baryons, the influence of chiral symmetry, or changes of the in-medium mass. Therefore we do not expect our simplified model to describe all details of the data. However, what we have shown with this comparison is that the self-consistent description of collisional broadening, which has been neglected in the analysis so far has a strong influence on the theoretical description of the spectra in the dilepton yield in the low-invariant-mass $200 \, {\rm MeV}<M<750 \, {\rm MeV}$ range.

\section{Conclusion and Outlook}\label{summary}

We have investigated the in-medium properties of $\rho$-mesons in
a self-consistent calculation taking into account finite-width effects due to 
the broadening induced by the $\pi$-$\rho$ interaction at finite temperature.
For this purpose we treat the $\rho$-mesons as a massive vector-mesons in a Stückelberg formalism and resum the $\rho-\pi$-interaction
considering finite-widths effects only.

Special emphasis was given to the fact that this resummation in the
Cornwall Jackiw Tomboulis formalism violates the Ward Takahashi identies connected to the Stückelberg field and therefore gauge invariance. To circumvent these difficulties we have worked in a gauge
which guarantees the transversality of the $\rho$-meson spectral-function in the resummation scheme.
We also employed another approach first introduced by van Hees and Knoll in \cite{vanHees:2000bp} where they discussed self-consistent finite width effects on the $\rho$ meson by pions in a similar framework. In that approach (approach II) one enforces the Ward Takahashi identities by disregarding the temporal components of the self-energy.

We derived in detail self-consistently coupled 
Schwinger Dyson eqns. in the finite width approximation for the $\rho$-meson and the pions for both approaches and 
reformulated these equations into coupled integral-equations giving explicit 
expressions for the imaginary parts of the self-energies and 
the spectral-functions.

Discussing the numerical solution of 
these equation gave physical insight into the 
broadening effects of the $\rho$-meson and the pion in a hot environment.

This study demonstrates that $\rho$-mesons embedded in a pure pion gas exhibit substantial collisional broadening, in contrast to perturbative approaches. This shows that a large broadening of the pions does not necessarily require a high density of baryons or chiral symmetry restoration.
 
In order to demonstrate that these effects are quantitatively 
important and contribute to the understanding of the experimentally investigated 
dilepton spectra we made a comparison of the results with data obtained by the CERES NA49 collaboration using a parametrized fireball evolution model.

We emphasize that this comparison does not aim at a description of all details of the data (where many important effects that we
disregard in our model (as scattering off by baryons, chiral symmetry and resulting
mass shifts) would have to be taken into account). 

We demonstrated that the investigated self-consistent finite-width effects on
the $\rho$-meson are an in-medium modification whose order of magnitude is not suppressed as compared to other effects. Therefore, they are relevant to the understanding of the experimentally oberserved enhanced dilepon yield (as compared to naive extrapolations of p-A collisions)  in the low invariant mass region $200 {\rm ~MeV} < M < 600 {\rm ~MeV}$ in A-A collision.

The phenomenological most important remaining questions might be how to handle and disentangle the additional effects of chiral symmetry restoration, 
interactions with baryons, and corresponding mass shifts
additionally to the finite width effects on the $\rho$ spectral function. 

Regarding the interaction with baryons, several studies already deal with the inclusion of medium modifications of vector mesons which arise from the modification of pionic modes modified through the nuclear medium
at finite density and temperature \cite{Urban:1998eg,Urban:1999im,Riek:2004kx}. Especially the last paper includes a treatment of the nucleon and the $\Delta(1232)$-isobar resonance in a $\Phi$-derivable approximation supplemented by a Migdal's short range correlation for the particle-hole excitation. 
 
From a theoretical point of view one might wish to include the handling of renormalization effects \cite{HendrikDiss,VanHees:2001pf,vanHees:2002bv} properly going
beyond the finite width approximation and improve in the direction 
of feasible gauge-invariant resummation schemes which could also be interesting 
for the study of fundamental theories like QCD beyond HTL approximation. 

\section*{Acknowledgement}
T. Renk and J. Ruppert acknowledge support by the Alexander 
von Humboldt-Foundation as Feodor-Lynen-Fellows. The Center of Scientific Computing (CSC) of 
J.W. Goethe University is acknowledged for providing computer access to its
cluster. 
We thank Steffen Bass, Marcus Bleicher, Adrian Dumitru, Dennis D. Dietrich, Stefan
Hofmann, Berndt M\"uller, Dirk H. Rischke, Dirk R\"oder, and J\"urgen Schaffner-Bielich for fruitful discussions. 

\appendix
\section{Projection operators for vector particles}
\label{projector}
In this section of the appendix, we introduce a projection
operator\footnote{N.b.: although the literature refers to the
introduced operators as "projection operators", one has always to
bear in mind that the do not project onto orthogonal subspaces.
This becomes obvious considering eqns. (\ref{tensorinv}).} formalism that is
suitable for the description of the propagating modes of vector
particles in the medium. The notation is similar to \cite{Rischke:2002rz} . 
In the
presence of a heat bath as a special frame of reference,
lorentz-invariance is broken by the medium. Therefore the
spectral densities become functions of the energy $p_0$ and the
magnitude of the three-momentum $|\bf{p}|$ of the particles with
four-momentum $P=(p_0,\bf{p})$ relative to the medium.

We introduce the projector ${\rm E}$ projecting onto the subspace
parallel to $P^\mu$:
\begin{equation} \label{E}
{\rm E}^{\mu \nu} \equiv \frac{P^\mu \, P^\nu}{P^2}\,\, .
\end{equation}
Then we introduce a vector $N^\mu$ which is orthogonal to
$P^\mu$:\begin{equation} N^\mu \equiv \left( \frac{p_0\,
p^2}{P^2}, \frac{p_0^2\, {\bf p}}{P^2} \right) = \left(g^{\mu \nu}
- {\rm E}^{\mu \nu}\right)\, f_\nu\,\, ,
\end{equation}
where $f^\mu=(0,\vec{p})$. Then $N^2=-p_0^2p^2/P^2$. The
definition of the other three projectors is:
\begin{equation} \label{BCA}
{\rm B}^{\mu \nu} \equiv \frac{N^\mu\, N^\nu}{N^2}\,\,\,\, ,
\,\,\,\,\, {\rm C}^{\mu \nu} \equiv N^\mu \, P^\nu + P^\mu\, N^\nu
\,\,\,\, , \,\,\,\,\, {\rm A}^{\mu \nu} \equiv g^{\mu \nu} - {\rm
B}^{\mu \nu} - {\rm E}^{\mu \nu} \,\, .
\end{equation}
The tensor ${\rm
A}^{\mu \nu}$ projects onto the spatial transverse subspace of
$P^\mu$:
\begin{equation}
{\rm A}^{00} = {\rm A}^{0i}=0\,\,\,\, , \,\,\,\,\, {\rm A}^{ij} =
- \left(\delta^{ij} - \hat{p}^i \, \hat{p}^j \right)\,\, .
\end{equation}

 The tensor ${\rm B}^{\mu \nu}$ projects onto the spatial longitudinal
subspace of $P^\mu$:
\begin{equation}
{\rm B}^{00} = - \frac{p^2}{P^2} \,\,\,\, , \,\,\,\,\, {\rm
B}^{0i} = - \frac{p_0\, p^i}{P^2}\,\,\,\, ,\,\,\,\, {\rm B}^{ij} =
- \frac{p_0^2}{P^2}\,\hat{p}^i\,\hat{p}^j\,\, .
\end{equation}
 This can be written as:\bea B^{\mu \nu}=g^{\mu
\nu}+\delta^{ij}-\frac{P^\mu P^\nu}{P^2}-\hat{p}^i \hat{p}^j\,\,,
\eea where for $\mu=0$ or $\nu=0$ the terms $\delta^{ij}$ and
$\hat{p}_i \hat{p}_j$ don't contribute, elsewise $i=\mu$ and
$j=\nu$. The projectors $\rm C$ and $\rm E$ have the following
forms, respectively: \bea {\rm C}^{00} &=&  2\frac{p_0^2 p^2}{P^2}
\,\,\,\, , \,\,\,\,\, {\rm C}^{0i} = \frac{p p_0
(p_0^2+p^2)}{P^2}\,\hat{p}^i\,\,\,\, ,\,\,\,\, {\rm C}^{ij} = 2
\frac{p_0^2 p^2}{P^2}\,\hat{p}^i\,\hat{p}^j\,\, , \\{\rm E}^{00}
&=& \frac{p_0^2}{P^2} \,\,\,\, , \,\,\,\,\, {\rm E}^{0i} =
\frac{p_0 p }{P^2}\,\hat{p}^i\,\,\, ,\,\,\,\, {\rm E}^{ij} =
\frac{p^2}{P^2}\,\hat{p}^i\,\hat{p}^j\,\, . \eea

The tensorial decomposition of self-energies, spectral functions,
propagators etc. (represented here by the generic variable $X$)
can be written as:\begin{equation} \label{tensor} X^{\mu \nu}(P) =
X^{\rm a}(P) \, {\rm A}^{\mu \nu} + X^{\rm b}(P) \, {\rm B}^{\mu
\nu} + X^{\rm c}(P)\, {\rm C}^{\mu \nu} + X^{\rm e}(P)\, {\rm
E}^{\mu \nu}\,\, .
\end{equation}

The scalar components of the projection $X^a$, $X^b$, $X^c$ and
$X^e$ can be obtained by projecting $X^{\mu \nu}$ onto these
components. Using the abbreviations:
\begin{equation}
X^t(P) \equiv \frac{1}{2} \, \left( \delta^{ij} - \hat{p}^i\,
\hat{p}^j \right) \, X^{ij}(P) \,\,\,\, ,\,\,\,\,\, X^\ell(P)
\equiv \hat{p}_i \, X^{ij}(P)\, \hat{p}_j \,\, ,
\end{equation} one obtains the following
equations: \begin{subequations} \label{XProj}
\begin{eqnarray}
X^{\rm a}(P) & = & \frac{1}{2}\, X^{\mu \nu}(P)\,
{\rm A}_{\mu \nu} =  -  X^t(P) \,\, ,  \label{Pia} \\ \label{Pib}
X^{\rm b}(P) & = & X^{\mu \nu}(P)\, {\rm B}_{\mu \nu} \nn \\ &=& -
\frac{p^2}{P^2} \, \left[ X^{00}(P) + 2\, \frac{p_0}{p}\,
X^{0i}(P)\,\hat{p}_i
+ \frac{p_0^2}{p^2}\, X^\ell(P) \right] \,\, ,  \\
X^{\rm c}(P) & = & \frac{1}{2\, N^2 \, P^2}\, X^{\mu \nu}(P)\,
{\rm C}_{\mu \nu} \,\, \\&=& -\frac{1}{P^2}\, \left[ X^{00}(P) +
\frac{p_0^2+p^2}{p_0\,p}\, X^{0i}(P)\,\hat{p}_i
+ X^\ell(P) \right] \,\, , \\
X^{\rm e}(P) & = & X^{\mu \nu}(P)\, {\rm E}_{\mu \nu}  \nn \\&=&
\frac{1}{P^2}\, \left[ p_0^2 \, X^{00}(P) + 2\,p_0\,p \, X^{0i}(P)
\, \hat{p}_i + p^2 \, X^\ell(P) \right] \,\, .
\end{eqnarray}
\end{subequations}
 If the tensor is four-dimensionally transverse
(e.g. $P_{\mu}X^{\mu \nu}=0$) these equations simplify
significantly:
\begin{subequations} \label{transTensor}
\begin{eqnarray}
X^{\rm a}(P) & = &  -  X^t(P) \,\, ,  \\
X^{\rm b}(P) & = &  -  \frac{P^2}{p_0^2} X^{\ell}(P) \,\, ,  \\
X^{00}(P) \,& = & \frac{p^2}{p_0^2} X^{\ell}(P) \,\, , \\
X^{0i}(P)\,\hat{p}_i & = & -\frac{p}{p_0} X^{\ell}(P) \,\, , \\
X^{\rm c}(P) & = & 0 \,\, \\
X^{\rm e}(P) & = & 0 \,\, .
\end{eqnarray}
\end{subequations}

The inverse of a tensor $X^{\mu \nu}$ (if it exists) is:
\begin{equation} \label{tensorinv}
X^{-1}_{\mu \nu}(P) = X^{-1, \, \,\rm a}(P) \, {\rm A}_{\mu \nu} +
X^{-1, \, \,\rm b}(P) \, {\rm B}_{\mu \nu} + X^{-1, \, \,\rm
c}(P)\, {\rm C}_{\mu \nu} + X^{-1, \, \,\rm e}(P)\, {\rm E}_{\mu
\nu}\,\, , \end{equation} where \bea \label{tensor2} X^{-1, \,
\,\rm a}(P)=\frac{1}{X^{\rm a}} \, \, &,& \, \,
 X^{-1, \, \,\rm b}(P)=\frac{X^{\rm e}}{\delta} \nn \\
  X^{-1, \, \,\rm c}(P)=-\frac{X^{\rm c}}{\delta} \, \, &,& \, \,    X^{-1, \, \,\rm e}(P)=\frac{X^{\rm b}}{\delta}
\eea and $\delta= X^{\rm b} X^{\rm e}-P^2N^2 \left(X^{\rm
c}\right)^2 $. One checks explicitly that indeed $X^{\alpha
\gamma}X^{-1}_{\gamma \beta} = g^{\alpha}_{\beta} $.

\section{Thermal dilepton production rate in a VMD model}\label{production}
In this section of the appendix, we quote some important formulas for 
the dilepton production rate via the decay of a thermal $\rho$-meson
in a VMD-model
We use the formalism as in McLerran
and Toimela \cite{McLerran:1985ay}. 
Relating the electromagnetic current-current correlator  \be
W^R_{\mu\nu}(q)\equiv i\int d^4x e^{iq\cdot x}\theta(x_0)\sum_H\,
\langle H\,\mid [J^h_{\mu}(x),J^h_{\nu}(0)]\mid\,H\,\rangle\,
\frac{e^{-\beta\,E_H}}{Z(\beta)}. \label{retcor} \ee to the photon
propagator via \bea W^R_{\mu \nu}(Q) = Q^2 g_{\mu \gamma} D_{{\rm
R}\,,{\rm Photo}}^{\gamma \delta} Q^2 g_{\delta \nu}=Q^4
D^{R}_{{\rm Photo }\,\mu \nu} \eea leads to the dilepton rate as a
functional of the current-current correlator \be
\frac{dR}{d^4q}=\frac{\alpha}{12\pi^4q^2}\left(1+\frac{2m_L^2}{q^2}\right)
\sqrt{1-\frac{4m_L^2}{q^2}}g_{\mu \nu} {\rm Im} W_{\rm R}^{\mu
\nu}f_{}(q_0) \label{drd4q02} \, \,.\ee These equation are
generally valid independent of the modelling of the
hadronic electromagnetic current.

In vector meson dominance (VMD) models is the retarded
current-current correlator proportional to the retarded $\rho$-meson
propagator (up to order $\frac{e}{g}$):\bea
\label{currentRhoProp} W^{\rm R}_{\mu \nu}=\frac{e^2}{g^2}m_{\rho }^4{\cal S}^{R}_{\mu \nu} \eea

The equation for the dilepton production rate from the hot
hadronic medium as a functional of the $\rho$-meson spectral-function
is now: \be \label{rateprod} \frac{dR}{d^4q}=-\frac{1}{6\pi^3}\frac{\alpha^2}{g^2}
\frac{m_{\rho }^4}{q^2} \left(1+\frac{2m_L^2}{q^2}\right)
\sqrt{1-\frac{4m_L^2}{q^2}}g_{\mu \nu} \rho_{\rm}^{\mu \nu} \,
f_{}(q_0) \, \,.\ee Here, $(dR/d^4q)$ is the
production rate of dileptons with a total invariant mass $q^2$ and
the rest mass $m_L$ of one lepton.

\bibliography{u4}
\end{document}